\documentclass[showpacs, pra,twocolumn,preprintnumbers ,amsmath, amssymb, superscriptaddress, aps]{revtex4-2}
\usepackage{pgfplots}
\usepackage{tikz}
\pgfplotsset{compat=1.18}
\usepackage{color}
\usepackage{amsmath,amssymb}
\usepackage{pifont}
\usepackage{amssymb}  
\usepackage{bbold}
\usepackage{float}
\usepackage{subfloat}

\usepackage[caption=false]{subfig}
\usepackage{tikz}
\usepackage{makecell}
\usepackage{subfig}
\usepackage{pifont}   
\usepackage{graphicx} 
\graphicspath{{Figures/}}
\usepackage{dcolumn}  
\usepackage{bm}       
\usepackage{multirow} 
\usepackage{placeins}
\usepackage[colorlinks]{hyperref}
\hypersetup{colorlinks=true,linkcolor=red}
\usepackage{mathtools}
\usepackage{appendix}

\def \be{\begin{align}}
	\def \ee{\end{align}}
\def \bea{\begin{eqnarray}}
	\def \eea{\end{eqnarray}}

\begin{document}
	
	\title{Energy-gap–controlled current oscillations in graphene under periodic driving}

	\author{Hasna Chnafa}
	\affiliation{Laboratory of Theoretical Physics, Faculty of Sciences, Choua\"ib Doukkali University, PO Box 20, 24000 El Jadida, Morocco}
	
	\author{Clarence Cortes}
	\affiliation{Vicerrector\'ia de Investigaci\'on y Postgrado, Universidad de La Serena, La Serena 1700000, Chile}  
	\author{David Laroze}
	\affiliation{Instituto de Alta Investigaci\'on, Universidad de Tarapac\'a, Casilla 7D, Arica, Chile}
	

			\author{Ahmed Jellal}
			\email{a.jellal@ucd.ac.ma}
			\affiliation{Laboratory of Theoretical Physics, Faculty of Sciences, Choua\"ib Doukkali University, PO Box 20, 24000 El Jadida, Morocco}
			
			\begin{abstract}				
We investigate the impact of an induced mass term $\Delta$ on the current density in graphene subjected to {a space- and time-dependent periodic potential $U(x,t)$. By solving the Dirac equation and deriving both the quasi-energy spectrum and the corresponding eigenspinors, we obtain explicit analytical expressions for the current density, which exhibits a clear dependence on $\Delta$.
We show that $\Delta$ acts as a tunable control parameter that governs the amplitude, sign, and resonance structure of Josephson-like current oscillations. For} normal incidence and a purely time-periodic potential, our results reveal that the oscillations within the energy gap gradually diminish as the mass term $\Delta$ increases. This suppression leads to a weakening of the Josephson-like effect typically observed in such systems. When the potential $U(x,t)$ is periodic in both space and time, the behavior becomes more complex. The current density can take either positive or negative values depending on the magnitude of the induced gap, and it generally decreases over time. As a result, the resonance phenomena—prominent at lower gap values—become progressively less significant as $\Delta$ increases. {These findings underscore the tunable nature of light-matter interactions and quantum transport in gapped graphene, suggesting potential applications in terahertz (THz) nanoelectronic devices and optically controlled quantum switches.}


				\pacs{72.80.Vp, 73.22.Pr, 73.40.Gk, 78.67.Wj\\
				{\sc Keywords}: Graphene, periodic potential, gap, Josephson current.}
			\end{abstract}
			
			\maketitle

	\section{Introduction}


	Graphene, a monolayer of carbon atoms arranged in a two-dimensional honeycomb lattice, exhibits remarkable electronic properties that have attracted intense interest from both theoretical and experimental communities \cite{1,2}. Its low-energy charge carriers behave as massless Dirac fermions, governed by a Dirac-like Hamiltonian \cite{7,8,9,10}. These quasi-particles travel with an effective Fermi velocity of approximately $10^6\, \text{m/s}$, mimicking relativistic particles in a solid-state environment and drawing analogies with neutrinos \cite{9}. The linear dispersion relation, with conduction and valence bands meeting at the Dirac points ($K$ and $K'$) in the Brillouin zone, makes graphene a zero-gap semiconductor or a semimetal \cite{1,2,11}.
	While this gapless nature underlies many of graphene’s exotic properties, it presents a major challenge for technological applications, especially in digital electronics. For instance, field-effect transistors (FETs) require a finite energy gap to achieve efficient switching between “ON” and “OFF” states. To make graphene compatible with logic applications, various approaches have been proposed to induce or engineer a tunable band gap.
	
	Experimentally, band gaps have been achieved by placing graphene on substrates that break its sublattice symmetry. For example, depositing graphene on hexagonal boron nitride (h-BN) can open a gap of about 100 meV \cite{12,13}, while growth on silicon carbide (SiC) leads to larger gaps up to 260 meV \cite{14,15}. Similarly, metallic substrates such as Cu(100) and Cu(111) can induce gaps around 250 meV \cite{16}. In bilayer graphene, applying a perpendicular electric field enables the creation of a tunable band gap, offering additional control \cite{17}.
	Recent advances have also explored more dynamic or structural routes for bandgap engineering. Floquet engineering—using periodic driving via light fields—has emerged as a promising method to induce topological and tunable gaps in graphene systems \cite{31,32,33}. 
	 Another approach, the formation of moiré superlattices in twisted bilayer and multilayer graphene allows for highly tunable and even correlated insulating or superconducting states through band flattening and symmetry breaking \cite{34,35}. In particular, the interplay between light-matter interaction and moiré geometry in twisted systems has been shown to give rise to topological flat bands and novel real-space degeneracies \cite{36,37}.
	Collectively, these techniques illustrate the ongoing efforts to overcome the intrinsic gaplessness of graphene, a critical step toward realizing graphene-based transistors, photodetectors, and quantum devices.


	The Josephson effect \cite{20,18} is a remarkable quantum phenomenon where a supercurrent—an electrical current with zero resistance—flows between two superconductors separated by a thin barrier made of an insulator or a normal (non-superconducting) metal. This current depends sinusoidally on the phase difference between the superconducting wavefunctions across the junction. The effect was first predicted in 1962 by British physicist Brian David Josephson, after whom it is named \cite{21}.
	Due to its inherently nonlinear nature, the Josephson effect gives rise to several fascinating phenomena. One well-known example is the appearance of Shapiro steps in the current-voltage (I–V) characteristics when the junction is exposed to electromagnetic radiation \cite{20}. These steps reflect the quantized nature of the phase locking between the superconducting current and the external drive.
	Graphene-based Josephson junctions have attracted growing attention in recent years. Studies have shown that the tunneling conductance in normal graphene/superconducting graphene (NG/SG) junctions \cite{22}, as well as in insulating graphene (IG) systems and more complex SG/IG/SG setups \cite{23,24}, exhibits oscillatory behavior as a function of the gate voltage applied to the barrier. This unique feature arises from the relativistic, massless Dirac fermions in graphene, whose behavior leads to interference patterns in the transmission of Cooper pairs through the junction \cite{24}.

	In addition to the well-known superconducting Josephson effect in superconducting junctions, recent studies have shown that periodically driven Dirac systems, such as graphene, can exhibit analogous oscillatory currents, often referred to as Josephson-like currents. These currents arise from coherent, non-equilibrium carrier dynamics under time-periodic potentials or electromagnetic fields~\cite{Sentef2015,Park2022,Broers2021,Ying2018}. 
	In this non-superconducting context, the oscillatory current originates from photon-assisted or Floquet-engineered transitions of Dirac fermions rather than Cooper-pair tunneling. This mechanism offers several advantages. It allows for the tunable control of current amplitude and phase via external drive parameters, such as field strength, frequency, and barrier geometry. Additionally, it operates at higher temperatures and eliminates the need for superconducting contacts. Furthermore, phase coherence is dynamically generated rather than being an equilibrium ground-state property. This allows for the fast modulation of transport and photon-driven interference effects. These features make optically driven Josephson-like currents a promising platform for exploring phase-coherent transport in two-dimensional materials and for developing novel optoelectronic devices.

	
{The above interesting properties of Josephson-like currents in optically driven Dirac systems have motivated further theoretical investigations. Previous works have addressed related aspects of periodically driven Dirac systems. In particular, Kibis \textit{et al.}~\cite{kibis} investigated the interaction of electrons in gapped Dirac materials with a strong off-resonant electromagnetic field (dressing) using Floquet theory. Their study mainly focused on the modification of the quasienergy spectrum and the effective Hamiltonian, demonstrating that the band gap and spin–orbit splitting can be significantly renormalized depending on the polarization of the dressing field.
	On the other hand, Savel'ev \textit{et al.}~\cite{25} developed a theoretical method to solve the massless Dirac–Weyl equation describing electron transport in graphene under a scalar potential barrier with arbitrary space- and time-dependence. In that work, a resonant enhancement of electron backscattering and oscillatory currents was predicted when the modulation frequency satisfies certain resonance conditions, analogous to the Shapiro steps in driven Josephson junctions. 
	However, these studies have largely focused on only one aspect: the electronic spectrum or transport in gapless graphene. As a result, the role of a finite mass term in Josephson-like current oscillations of periodically driven graphene remains poorly understood.


Motivated by this open question, we examine how the mass term $\Delta$ influences Josephson-like currents in graphene subjected to a space- and time-dependent periodic potential $U(x,t)$ applied along the $x$ direction. Two different forms of the potential are considered, and the Dirac equation is solved to determine the corresponding energy spectrum. Using the obtained eigenspinors, we derive the associated current density and analyze its dependence on the mass term $\Delta$.
We then perform a numerical study to explore how the Josephson-like current oscillations and their resonance features evolve under different conditions. The study demonstrates that current oscillations decrease with increasing mass gap when energy exceeds the mass gap at normal incidence and time-periodic potential conditions.
For finite transverse momentum ($k_y \neq 0$), the mass term $\Delta$ introduces additional modifications that significantly influence the current behavior. When the potential is periodic in both space and time, the current displays aperiodic oscillations whose sign and magnitude depend on $\Delta$. In this situation, the oscillations progressively decay over time, suggesting a weakening of the resonance response.
Our findings suggest that the current density can be adjusted by modifying the mass term $\Delta$,  which could lead to strategies for controlling Josephson-like transport in graphene-based systems.

}

	
	The manuscript is organized as follows. In Sec. \ref{II}, we introduce the theoretical model for gapped graphene subjected to a space- and time-dependent periodic potential, and we solve the Dirac equation to obtain the eigenspinors. Sec. \ref{III} focuses on the first type of potential, where we calculate the corresponding current density. We also draw an analogy to Shapiro steps and perform numerical analysis to explore how the current behaves under different physical parameters. In Sec. \ref{IV}, we extend our study to a second potential type, which varies both spatially and temporally. This allows us to examine the key features of the system in more detail, with special attention to how the mass term influences the results. {Sec. \ref{V} discusses the regime of validity of our model and experimental considerations for observing the predicted Josephson-like currents}. Finally, we summarize and conclude our findings.
	
	\section{Theoretical Model} \label{II}
	
	We consider Dirac fermions in gapped graphene   subjected to a space- and time-dependent scalar potential \( U(x,t) \) applied along the \( x \)-direction. The system is described by the following Hamiltonian
	\begin{equation}\label{hj}
		H=v_{F}[{\tau}_{z}{\sigma}_{x} {p}_{x}+{\sigma}_{y}{p}_{y}]+U(x,t)\mathbb{I}+\Delta {\sigma}_{z} 
	\end{equation}
	\\
	where \( v_{F} = 10^{6} \, \text{m/s} \) is the Fermi velocity, \( \tau_{z} = \pm 1 \) denotes the valley index, \( \sigma_{i} \) are the Pauli matrices, \( \mathbb{I} \) is the \( 2 \times 2 \) identity matrix, and \( \Delta \) represents the energy gap. 
	Due to translational invariance along the \( y \)-direction, the spinor can be written as \( \Phi(x, y, t) = \Psi(x, t) e^{i k_{y} y} \). Substituting this form into the Dirac equation, we obtain
%
	\begin{widetext}
	\begin{equation}\label{mp}
	\begin{pmatrix}
			U(x,t)& -i\hbar v_{F}\tau _{z}\partial_{x}\\
			-i\hbar v_{F}\tau _{z}\partial_{x}&U(x,t) \\
		\end{pmatrix}
		\Psi + i\hbar v_{F}k_{y}
		\begin{pmatrix}
			0 & -1\\
			1 & 0 \\
		\end{pmatrix}
		\Psi + \Delta 
		\begin{pmatrix}
			1 & 0\\
			0 & -1 \\
	\end{pmatrix}
	\Psi =i\hbar \partial_{t}\Psi.
	\end{equation}
	\end{widetext}
To obtain the solutions, we consider the case where the potential is switched on at positive times, namely \( U(x,t<0) = 0 \). We now introduce the following ansatz
	\begin{align}\label{mp2}
		&\Psi=\sum _{n=0}^{\infty}\left( i k_{y}\right)^{n}
			\begin{pmatrix}
			1 \\
			\tau_{z}  \\
		\end{pmatrix}
		\Psi_{n,+}+\sum _{n=0}^{\infty}\left(  i k_{y}\right)^{n}
			\begin{pmatrix}
			1 \\
			-\tau_{z}  \\
	\end{pmatrix}
	\Psi_{n,-}
	\end{align}
	where \( \Psi_{n,+} \) and \( \Psi_{n,-} \) represent the right- and left-propagating wave functions along the \( x \)-direction.
	{Note that the convergence of this expansion requires that the coefficients $\Psi_{n,\pm}$ remain bounded and do not grow faster than geometrically with the index $n$. This condition ensures a finite radius of convergence in $k_y$ and uniform validity of the expansion over time.} 
	Inserting~(\ref{mp2}) into~(\ref{mp}), we get
	\begin{align}\label{eq8}
		\left[U \mp i \hbar v_{F}\partial_{x} -i\hbar\partial_{t}\right]\Psi_{n,\pm}\pm\tau _{z}\hbar v_{F}\Psi_{n-1,\mp}+\Delta \Psi_{n,\mp}=0
	\end{align}
	with the condition \( \Psi_{-1,\pm}(x,t \geq 0) = 0 \) and \( U = U(x,t) \). After performing the rescaling \( \varepsilon = E / \hbar v_{F} \), \( \delta = \Delta / \hbar v_{F} \), and considering the case \( U(x,t) = 0 \), we find that the eigenspinor solutions of~(\ref{mp}) take the form
	\begin{widetext} 
	\begin{align} \label{R14}\Psi(x,t=0)= \alpha_{+}
		\begin{pmatrix}{1}\\
			{	s\sqrt{\left|\frac{\varepsilon-\delta}{\varepsilon+\delta}\right|}e^{i\theta}} 
		\end{pmatrix}e^{ik_{x}x}
		+\alpha_{-}
		\begin{pmatrix}
			1 \\
			-s\sqrt{\left|\frac{\varepsilon-\delta}{\varepsilon+\delta}\right|}e^{-i\theta}
		\end{pmatrix}e^{-ik_{x}x}
	\end{align}
	\end{widetext}
	where \( e^{\pm i\theta} = \dfrac{\tau_{z} k_{x} \pm i k_{y}}{\sqrt{k_{x}^{2} + k_{y}^{2}}} \), \( s = \text{sign}(\varepsilon + \delta) \), and \( \alpha_{\pm} \) are two constants. 
	To proceed further, we choose the initial states at \( t = 0 \) and set \( k_{y} = 0 \), which simplifies the expressions and corresponds to the case of normal incidence. Under these conditions, the spinor components become
	\begin{align}\label{eq555}
		\Psi_{0,\pm}(x) &= a_{0,\pm}(x) = {\alpha_{\pm}} \sqrt{|\varepsilon \pm \delta|} \, e^{\pm i k_{x} x} \\
		\Psi_{n>0,\pm} &= 0.
	\end{align}
%
%

\section{Temporally Periodic Potential} \label{III}

To begin our analysis, we fix the form of the potential \( U(x,t) \). As a first step, we consider the case where the potential is periodic in time but {linear} in space. This configuration allows us to isolate and understand the effects of temporal modulation on the dynamics of Dirac fermions in gapped graphene. We examine how this time dependence influences the energy spectrum and current density, particularly under the conditions of normal incidence.


	\subsection{Eigenspinors}
	
	The partial differential equation~(\ref{eq8}) can be solved using the method of characteristics~\cite{27}. Specifically, 
{we consider the case 
	\(
	U(x,t) = \frac{U_0}{L}x \sin(\omega t),
	\)
	which represents a potential that varies linearly along the \(x\)-direction and oscillates 
	sinusoidally in time. Although it is not spatially uniform, this form effectively describes 
	a homogeneous, time-periodic electric field 
	\(
	E(x,t) = -\partial_x U(x,t) = -\frac{U_0}{L}\sin(\omega t)
	\)
	acting on the Dirac fermions. 
	}
%
%
	Under these conditions, we find that the solutions to the equation are given by
	\begin{align}\label{eq16}
		\Psi_{n,\pm}=a_{n,\pm}(x,t)e^{-i\frac{U_{0}}{ L \hbar \omega }\left[(x-x\cos\omega t)\pm v_{F}(\frac{\sin\omega t}{\omega}-t)\right]}
	\end{align}
such that 
\begin{widetext}
		\begin{align}\label{eq14}
	&	a_{0,\pm}=a_{\pm}(x\mp v_{F} t)={\alpha_{\pm}}\sqrt{{\left|{\varepsilon\pm\delta}\right|}}e^{\pm ik_{x}(x\mp v_{F} t)}\\
	&\label{pm1}
	a_{n>0,\pm}=\mp i  v_{F}\tau_{z}  \int^{t}_{0} dt'\Psi_{n-1,\mp}(x\mp v_{F} t\pm v_{F}t',t')e^{i\frac{U_{0}}{ L \hbar \omega }\left[(x-x\cos\omega t')\pm v_{F}(t\cos\omega t'-t)\pm v_{F}\left(\frac{\sin\omega t' }{\omega}-t'\cos\omega t'\right)\right]}.
	\end{align}
\end{widetext}
%
%
	To zeroth-order approximation with respect to \( k_{y} \), i.e., by setting \( n = 0 \) in~(\ref{mp2}), we obtain the wave functions \( \Psi_{0,+} \) and \( \Psi_{0,-} \) from~(\ref{eq16}) and (\ref{eq14}), respectively. Finally, the general expression for the wave functions is given by
%
	\begin{widetext}
	\begin{align}\label{eq13}
		\Psi(x,t)=&\binom{1}{	\tau_{z}}
		{\alpha_+}\sqrt{{\left|{\varepsilon+\delta}\right|}}e^{+ik_{x}(x-v_{F}t)} e^{-i\frac{U_{0}}{L\hbar  \omega}\left[(x-x\cos\omega t)+v_{F}(\frac{\sin\omega t}{\omega}-t)\right]}\\
		&+ 
		\binom{1}{-	\tau_{z}}
		{\alpha_-}\sqrt{{\left|{\varepsilon-\delta}\right|}}e^{- ik_{x}(x+v_{F}t)}e^{-i\frac{U_{0}}{L\hbar  \omega}\left[(x-x\cos\omega t)-v_{F}(\frac{\sin\omega t}{\omega}-t)\right]}\nonumber,
	\end{align}
	\end{widetext}
	{which corresponds to the zeroth-order $k_y=0$ contribution and is well-behaved for all times, thus satisfying the convergence requirement at this order.}
	In particular, if the wave packet is initially purely right-moving, i.e., \( a_{-} = 0 \), then according to~(\ref{eq13}), it continues to propagate to the right for \( t > 0 \), and thus \( \Psi_{-} = 0 \). The first-order corrections with respect to \( k_{y} \) in (\ref{pm1}) can then be written as
%
	\begin{widetext}
	\begin{align}\label{eq18}
		a_{1,\pm}&=\mp iv_{F}{\tau_{z}}{\alpha_\pm}\sqrt{{\left|{\varepsilon\mp\delta}\right|}}e^{\mp ik_{x}(x\mp v_{F} t)}\int^{t}_{0} dt'e^{-2i v_{F}k_{x} t'}e^{\pm 2i\frac{U_{0}v_{F}}{L \hbar \omega}\left[\frac{\sin\omega t'}{\omega}-t'\right]}
	\end{align}
	\end{widetext}
	which can be written in terms of the Bessel function of the first kind as
	\begin{widetext}
	\begin{align}\label{ref1}
		a_{1,\pm}&=\mp iv_{F}\color{black}{\tau_{z}} {\alpha_\pm} \sqrt{{\left|{\varepsilon\mp\delta}\right|}}e^{\mp ik_{x}(x\mp v_{F}t)}\sum _{n=-\infty}^{+\infty}J_{n}\left(\frac{2U_{0}v_{F} }{L\hbar {\omega}^{2}}\right)\frac{e^{i\left(\mp\frac{ 2U_{0}v_{F}}{L\hbar\omega}-2 v_{F} k_{x}\pm n\omega\right)t}-1}{i\left(\mp\frac{ 2U_{0} v_{F}}{L\hbar\omega}-2 v_{F}k_{x}\pm n\omega \right)}.
	\end{align}
	\end{widetext}
	Now, we consider (\ref{mp2}), which represents the right- and left-propagating wave functions, and include the first-order corrections in \( k_{y} \) to obtain
	\begin{equation}\label{kh1}
		\Psi=\Psi_{+}
		\begin{pmatrix}
			1 \\
			\tau_{z}  \\
		\end{pmatrix}
		+\Psi_{-}
		\begin{pmatrix}
			1 \\
			-\tau_{z}  \\
	\end{pmatrix}
	\end{equation}
	where  the left and right functions are
	\begin{widetext}
	\begin{align}\label{eq20}
		\Psi_{\pm} =&{\alpha_\pm}\left[\sqrt{{\left|{\varepsilon\pm\delta}\right|}}e^{\pm ik_{x}(x\mp v_{F}t)}\pm k_{y}v_{F}\color{black}{\tau_{z}} \sqrt{{\left|{\varepsilon\mp\delta}\right|}}e^{\mp ik_{x}(x\mp v_{F}t)}\sum _{n=-\infty}^{+\infty}J_{n}\left(\frac{2U_{0}v_{F} }{{L \hbar  \omega}^{2} }\right)\frac{e^{i\left(\mp\frac{2U_{0} v_{F}}{L \hbar  \omega }-2 v_{F}k_{x}\pm n\omega\right)t}-1}{i\left(\mp\frac{2U_{0} v_{F}} {L \hbar  \omega }-2v_{F}k_{x}\pm n\omega \right)}\right]
	\nonumber	\\
		&
		e^{-i\frac{U_{0}}{L\hbar \omega}\left[(x-x\cos\omega t)\pm v_{F}(\frac{\sin\omega t}{\omega}-t)\right]}.
	\end{align}
	\end{widetext}
	{It includes the first-order correction in $k_y$ and remains bounded provided that the resonance condition
	$	
			\mp\frac{2U_0 v_F}{L\hbar\omega}-2v_F k_x\pm n\omega \neq 0
	$
		is fulfilled for all integers $n$. Under these non-resonant conditions, the series expansion remains convergent for all times. }
	

We now discuss the Josephson-like currents in our system by applying the results in the above section. Using the resulting wave functions and adding the first-order corrections, we investigated the current density due to the time-periodic potential. 
This approach allows us to gain insight into how the interplay between three key physical ingredients—the energy gap introduced by the mass term $\Delta$, the external time-dependent potential, and the transverse wave number $k_y$—influences charge transport in gapped graphene.


	
	\subsection{Josephson-like currents}
	

	We now determine the components of the current density in terms of the right- and left-propagating wave functions, \( \Psi_{+} \) and \( \Psi_{-} \). These components are given by
	\begin{align}\label{eq22}
		&j_{x}
		=2v_{F}\left(\Psi^{\ast}_{+}\Psi_{+}-\Psi^{\ast}_{-}\Psi_{-}\right)=j_{0x}+j_{1x} \\\label{eq45}
		&j_{y}
		=2i v_{F}\tau_{z}\left(\Psi^{\ast}_{+}\Psi_{-}-\Psi_{+}\Psi^{\ast}_{-}\right)=j_{0y}+j_{1y}.
	\end{align}
	{In the following analysis, we set $\alpha_+= \alpha_-=\alpha$ to simplify the calculation of physical quantities.
		We start by examining the \( j_x \) component of the current density. The zeroth-order term gives the main contribution, while the first-order term introduces a correction arising from the transverse momentum. These terms are expressed, respectively, as}
		\begin{align}\label{eq25}
		j_{0x} =& 2 \alpha^{2} v_{F}\left({\left|{\varepsilon+\delta}\right|\left|e^{ik_{x}(x-v_{F}t)}\right|^{2}}-{\left|{\varepsilon-\delta}\right|\left|e^{-ik_{x}(x+v_{F}t)}\right|^{2}}\right)	\end{align}
	\begin{widetext}
	\begin{align}
	 \label{eq27}
		j_{1x} =& 
		4v_{F}^{2}k_{y}\tau_{z}\Re e\left[\alpha^{2}\sqrt{{\left|{\varepsilon^{2}-\delta^{2}}\right|}} e^{2ik_{x}(x- v_{F}t)}\sum _{n=-\infty}^{+\infty}J_{n}\left(\frac{2U_{0}v_{F} }{L\hbar
			{\omega}^{2}}\right)\frac{e^{i\left(\frac{2U_{0}v_{F}}{L \hbar  \omega}+2v_{F}k_{x}-n\omega\right)t}-1}{i\left(\frac{2U_{0}v_{F}} {L \hbar  \omega}+2v_{F}k_{x}-n\omega \right)}\right]\\ \nonumber
		&+ 4v_{F}^{2}k_{y}\tau_{z}\Re e\left[\alpha^{2}\sqrt{{\left|{\varepsilon^{2}-\delta^{2}}\right|}} e^{-2ik_{x}(x+ v_{F}t)}\sum _{n=-\infty}^{+\infty}J_{n}\left(\frac{2U_{0} v_{F} }{L \hbar  {\omega}^{2}}\right)\frac{e^{i\left(-\frac{2U_{0}v_{F}}{L \hbar  \omega}+2v_{F}k_{x}+n\omega\right)t}-1}{i\left(-\frac{2U_{0}v_{F}}{L\hbar\omega}+2v_{F}k_{x}+n\omega \right)}\right]. 
	\end{align}
	\end{widetext}
	
	{At this stage, a few remarks are in order.
		Expression~(\ref{eq25}) represents a valley-unpolarized quantity, as it shows no dependence on the valley index $( \tau_{z} )$, similar to the formulation in \cite{25}. However, unlike the result reported in \cite{25}, this expression does not vanish.
	}
	For an initially purely right-moving wave packet (\( a_{0,-} = 0 \)), we obtain \( j_{0x} = 2\alpha^{2}v_{F} \left|{\varepsilon+\delta}\right|\left|e^{ik_{x}(x-v_{F}t)}\right|^{2} \), which confirms the occurrence of perfect Klein tunneling through the potential \( U(x,t) \), regardless of its space-time dependence. This behavior is consistent with previous findings in \cite{25}. Furthermore, for \( k_{y} \neq 0 \), the first-order correction \( j_{1x} \) becomes relevant and may be exploited to engineer valley-polarized currents. Next, we derive the expression for \( j_{0y} \), that is
	\begin{equation}\label{eq28}
		j_{0y}=4v_{F}\tau_{z}\alpha^{2} \sqrt{\left|{\varepsilon^{2}-\delta^{2}}\right|}
		\sin 2\left[k_{x}x+\frac{U_{0}v_{F}}{L\hbar \omega}\left(t-{\frac{\sin\omega t}{\omega}}\right)\right]
	\end{equation}
	which can be written in terms of Bessel function as
	\begin{widetext}
	\begin{align}\label{kju1}
		j_{0y}=4v_{F}\tau_{z}\alpha^{2}
		\sqrt{\left|{\varepsilon^{2}-\delta^{2}}\right|}\sum _{n=-\infty}^{+\infty}J_{n}\left(\frac{2U_{0}v_{F} }{L \hbar  {\omega}^{2}}\right)\sin\left[2k_{x}x+\frac{2 U_{0}v_{F}}{L \hbar \omega}t-n{\omega}t\right]
	\end{align}
	\end{widetext}
	This behavior is reminiscent of the Shapiro steps in irradiated Josephson junctions, which appear when the frequency satisfies the resonance condition \cite{27}
	\begin{align}\label{kju}
		\omega_{n}=\sqrt{\frac{2U_{0}v_{F}}{L\hbar n
		}}=\omega, \quad  n\in\mathbb{N}.
	\end{align}
	{Finally, in the case \( k_{y} \neq 0 \), to compute the first-order correction \( j_{1y} \), we apply the initial conditions  
		\( a_{0,+} = \alpha \sqrt{|\varepsilon + \delta|}\, e^{i k_x (x - v_F t)} \) and \( a_{0,-} = 0 \)  
		to obtain
	}
	\begin{widetext}
	\begin{align}\label{eq29}
		j_{1y}=&-4 v_{F}^{2}{k_{y}}\alpha^{2} \left|{\varepsilon+\delta} \right|\sqrt{J_{n}^{2}\left(\frac{2U_{0} v_{F} }{{L\hbar \omega}^{2}}\right)\frac{\sin^{2}{\left(-\frac{U_{0}v_{F}}{L \hbar \omega}+v_{F}k_{x}+\frac{n\omega}{2}\right)t}}{\left(-\frac{U_{0}v_{F}} {L \hbar \omega}+v_{F}k_{x}+ \frac{n\omega}{2} \right)^{2}}} 
		\\
		&
		\sin\left[\frac{ 2U_{0}v_{F}}{L\hbar \omega}\left(t-\frac{\sin\omega t}{\omega}\right)+\arg\left({\alpha^{2}\left|{\varepsilon+\delta} \right|}e^{-2iv_{F}k_{x}t} \sum _{n=-\infty}^{+\infty}J_{n}\left(\frac{2U_{0} v_{F}}{L \hbar {\omega}^{2}}\right)\frac{e^{i\left(- \frac{2U_{0}v_{F}}{L \hbar \omega}+2v_{F}k_{x}+ n\omega\right)t}-1}{i\left(-\frac{2U_{0}v_{F}} {L \hbar\omega}+2v_{F}k_{x}+ n\omega \right)}\right)\right]. \nonumber
	\end{align}
	\end{widetext}
	We observe that the current component \( j_{0y} \) (\ref{eq28})
	{explicitly depends on the valley index \( \tau_{z} \) and is evaluated} 
	at normal incidence (\( k_{y} = 0 \)). 
	{The sinusoidal dependence observed in~(\ref{eq28}) is reminiscent of Josephson-like oscillations, which arise from the phase-dependent nature of the current.} 
	Accordingly, the total current can be expressed as 
	\( j^{\tau_{z}} = j^{+} + j^{-} \), where \( j^{\tau_{z}} \) denotes the 
	contributions from the \( K \) (\( \tau_{z} = +1 \)) and \( K' \) 
	(\( \tau_{z} = -1 \)) valleys.
%
	In contrast, the current correction \( j_{1y} \) (\ref{eq29}), depends solely on the transverse momentum \( k_{y} \), and captures the influence of small but finite incident angles (\( k_{y} \neq 0 \)) on the current density. Importantly, (\ref{eq29}) is derived under the resonance condition corresponding to Shapiro steps
	\begin{align}\label{k1ju}
		k_{n} = k_{x} = -\frac{n\omega}{2v_{F}} + \frac{U_{0}}{L\hbar\omega}, \qquad n \in \mathbb{N}
	\end{align}
	which governs the directional dependence of the wave vector \( k \)~\cite{25}.
	In what follows, we explore the behavior of the system under various configurations of physical parameters, with particular emphasis on the role of the mass gap \( \Delta \) in modulating the current densities.
	{We emphasis that  
when the resonance condition
	$
	-\frac{2U_{0}v_{F}}{L\hbar\omega} + 2v_{F}k_{x} + n\omega = 0
	$
	is fulfilled, the first-order contribution to the transverse current \(j_{1y}\) \eqref{eq29} develops a secular term proportional to time, leading to the gradual growth of the oscillation amplitude observed in Figs.~\ref{hde4}(a,b). This indicates that, at resonance, higher-order effects become relevant at sufficiently long times. For the parameter regime considered here, the first-order approximation provides an accurate description within the shown time range, but a more complete treatment including higher-order terms or a nonperturbative (e.g., Floquet) approach would be required to describe the long-time behavior beyond this regime.
	}

	\subsection{Numerical results}
	
\begin{figure}[H]\centering 
 	\subfloat[: ${U_{0}}v_{F}/L\hbar{\omega^{2}}=1/2$]	{\includegraphics[width=0.475\linewidth, height=0.09\textheight]{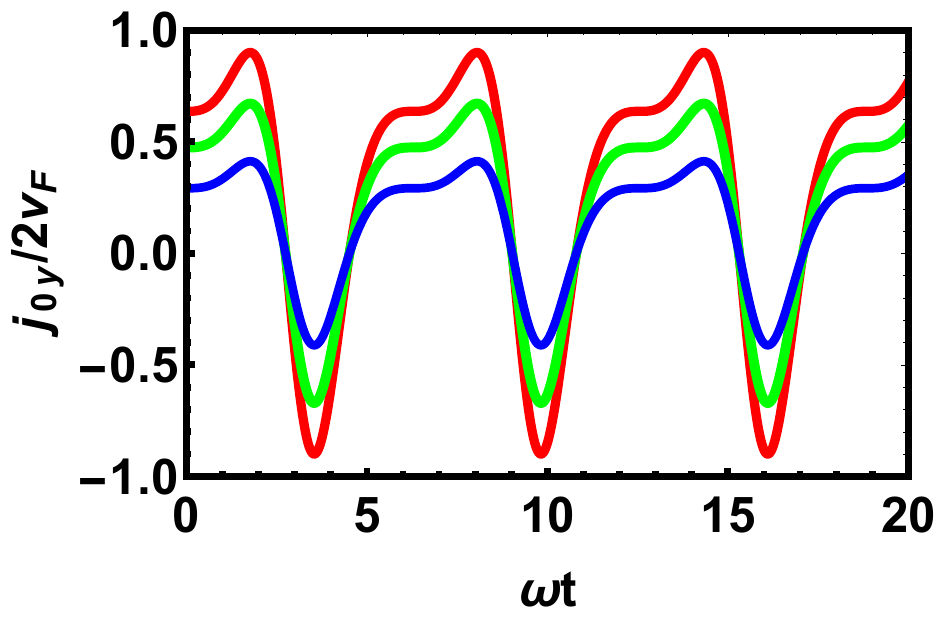}	\label{figure1}}
 	\subfloat[: ${U_{0}}v_{F}/L\hbar{\omega^{2}}=1$]	{\includegraphics[width=0.475\linewidth, height=0.09\textheight]{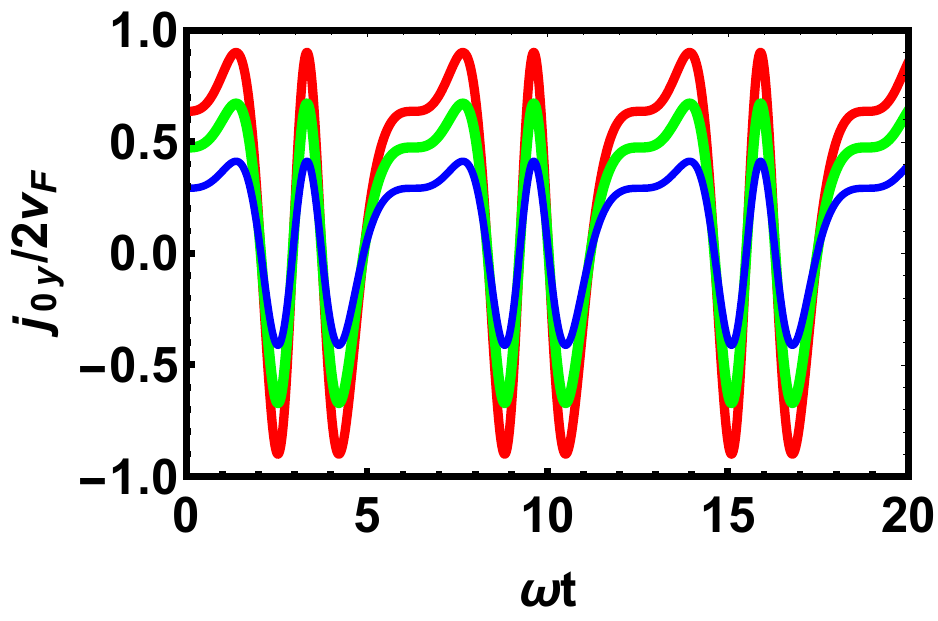}	\label{figureb}
 	}
 	\caption{Current density $j_{0y}/2v_{F}$ versus time $\omega{t}$ for $\tau_{z}=1$, ${k_{x}}x=\pi/8$, $\varepsilon L=0.9$, $\alpha=\sqrt{{L}/{2}}$, $L=1$ $\text{nm}$, $\delta L=0$ (red line), $0.6$ (green line), $0.8$ (blue line).  (a):  ${U_{0}}v_{F}/L\hbar{\omega^{2}}=1/2$,  (b):  ${U_{0}}v_{F}/L\hbar{\omega^{2}}=1$.}
 	\label{fe}
 \end{figure}

	To better understand the influence of gap on the derived current densities associated to a temporally periodic potential, perform a numerical analysis of the current density $j_{0y}$ ($\ref{eq28}$) at normal incidence $k_{y}=0$, assuming \( \varepsilon L > \delta L \) and considering a valley-polarized system (\( \tau_{z} = 1 \)).
	This analysis is conducted by varying key physical parameters.
 Fig.~$\ref{fe}$ presents the normalized current density $j_{0y}/2v_{F}$ versus time for ${k_{x}}x=\pi/8$, $\varepsilon L=0.9$, $\alpha=\sqrt{{L}/{2}}$, $L=1$ $\text{nm}$. We compare three different values of the rescaled gap  $\delta L=0$ (red line), $0.6$ (green line), $0.8$ (blue line). 
 According to the Shapiro steps condition (\ref{kju}),  $j_{0y}/2v_{F}$ oscillates in the range from positive to negative values. For \( \delta L = 0 \) (red), the oscillations cover \( -0.98 \leq j_{0y}/2v_{F} \leq 0.98 \), which agrees with  literature  \cite{25}. However, as the gap is growing the amplitude of these oscillations fall sharply indicating that the Josephson current starts to be suppressed.
 In Fig. \ref{fe}(b) for \( {U_{0}}v_{F}/L\hbar \omega^{2} = 1 \), we find that small amplitude oscillations (peaks) appear on top of the main periodicity in Fig. \ref{fe}(a) for {\color{black}\( {U_{0}}v_{F}/L\hbar \omega^{2} = 1/2 \)}. These extra peaks become more apparent and frequent in proportion as {\color{black}\( {U_{0}}v_{F}/L\hbar \omega^{2} \)} is large.
 Based on these results, we conclude that a larger gap $\delta$ significantly suppresses the Josephson current, while stronger driving amplitudes introduce higher-order resonances that further modulate the current response.

\begin{figure}[H]\centering 
	\subfloat[: ${U_{0}}v_{F}/L\hbar{\omega^{2}}=1/\pi$]	{\includegraphics[width=0.475\linewidth, height=0.09\textheight]{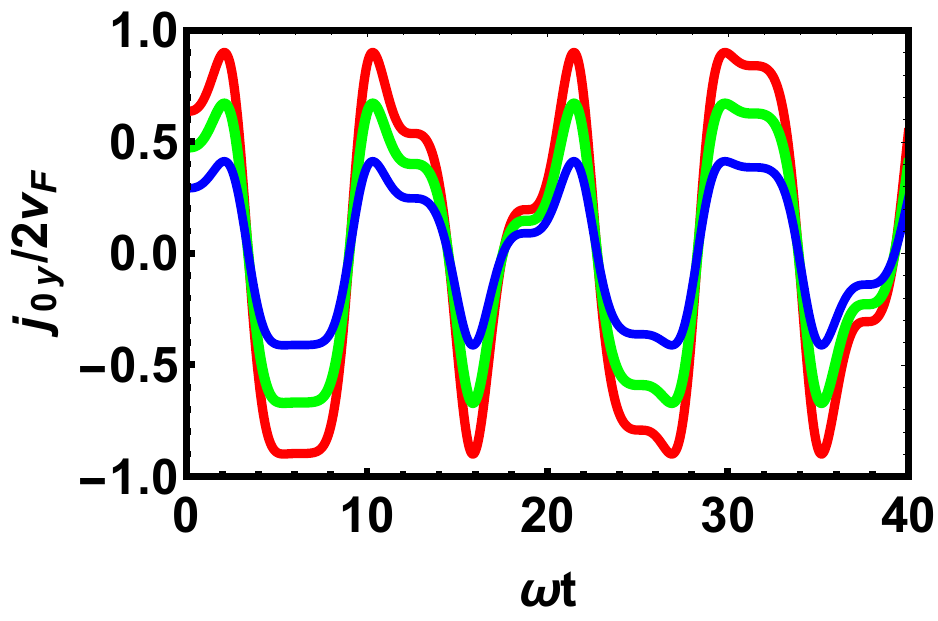}
		\label{figure5a1}}
	\subfloat[: ${U_{0}}v_{F}/L\hbar{\omega^{2}}=2/\pi$]	{\includegraphics[width=0.475\linewidth, height=0.09\textheight]{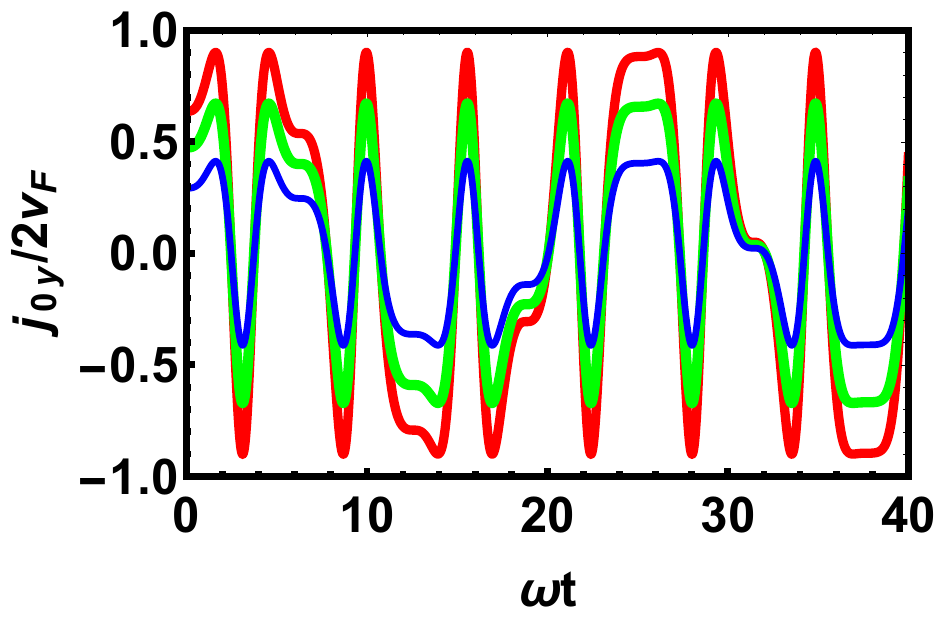}
		\label{figure6a1}}
	\caption{{Current density $j_{0y}/2v_{F}$ versus time  $\omega{t}$ for $\tau_{z}=1$, ${k_{x}}x=\pi/8$, $\varepsilon L=0.9$, $\alpha=\sqrt{{L}/{2}}$, $L=1$ $\text{nm}$, $\delta L=0$ (red line), $0.6$ (green line), $0.8$ (blue line).   (a):   ${U_{0}}v_{F}/L\hbar{\omega^{2}}=1/\pi$, 
			(b):    ${U_{0}}v_{F}/L\hbar{\omega^{2}}=2/\pi$.}}
	\label{ft1}
\end{figure}

Fig.~\ref{ft1} shows how the normalized transverse current density \( j_{0y}/2v_{F} \) evolves over time when the system is slightly off the Shapiro resonance defined in~(\ref{kju}). We examine two different driving conditions: \( {U_{0}}v_{F}/(L\hbar \omega^{2}) = 1/\pi \) and \( 2/\pi \). 
We also consider $\tau_{z}=1$, ${k_{x}}x=\pi/8$, $\varepsilon L=0.9$, $\alpha=\sqrt{{L}/{2}}$, $L=1$ $\text{nm}$ and three values of gap $\delta L=0$ (red line), $0.6$ (green line), $0.8$ (blue line). Note that for ($\delta L=0$), our results are similar to those obtained in \cite{25}. One observes that the aperiodic oscillations decrease with time as long as $\delta L$ increases. We also notice that the width of the oscillations, which located in regime $0\leq \omega{t} \leq 16$ of Fig.~{\ref{ft1}}(a),  becomes very small for the value ${U_{0}}v_{F}/L\hbar{\omega^{2}}=2/\pi$ as shown in Fig.~{\ref{ft1}}(b). As time increases, different oscillations appear. 
The results indicate that the current density is highly tunable via modifications of both the energy gap $\delta$ and the external potential amplitude $U_0$.

\begin{figure}[H]\centering
	\subfloat[: $\delta L=0$]
	{\includegraphics[width=0.5\linewidth, height=0.075\textheight]{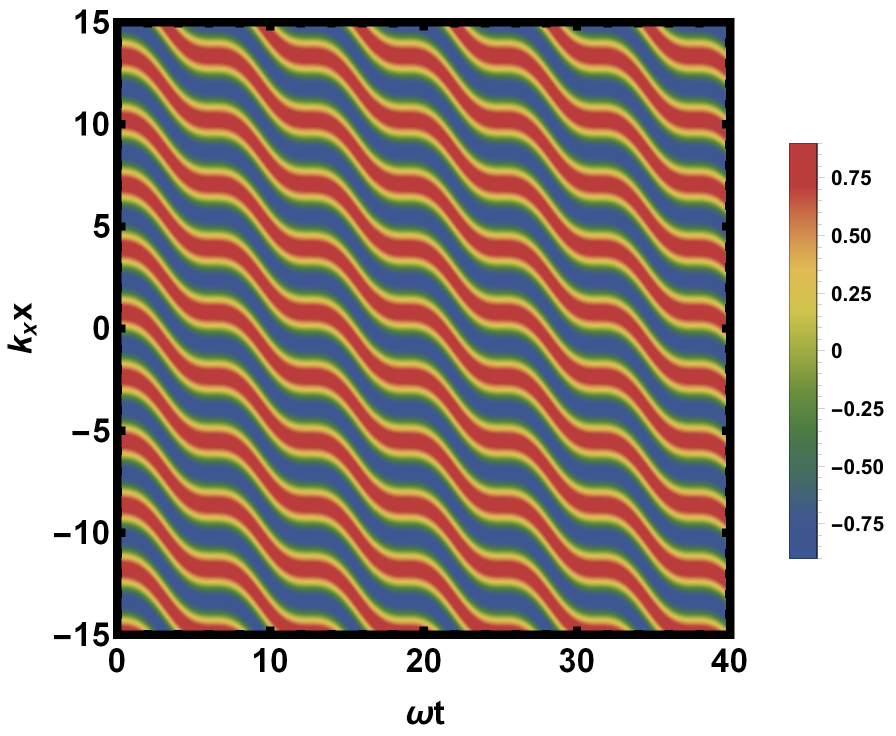}
		\label{fi}}
	\subfloat[: $\delta L=0.8$]
	{\includegraphics[width=0.5\linewidth, height=0.075\textheight]{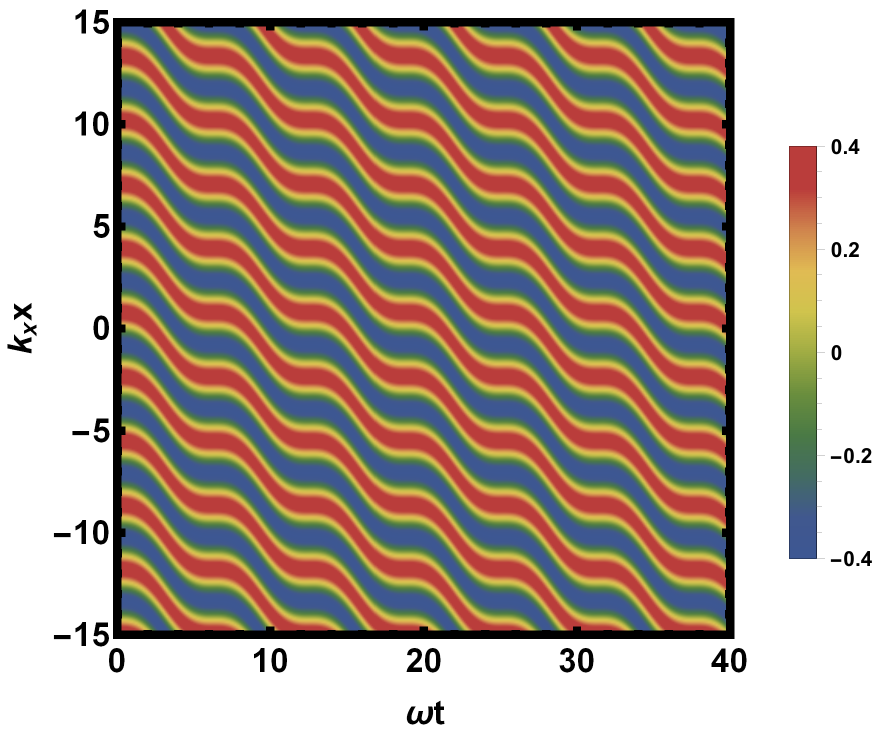}
		\label{fig40}}\\
	\subfloat[: $\delta L=0$]
{\includegraphics[width=0.5\linewidth, height=0.075\textheight]{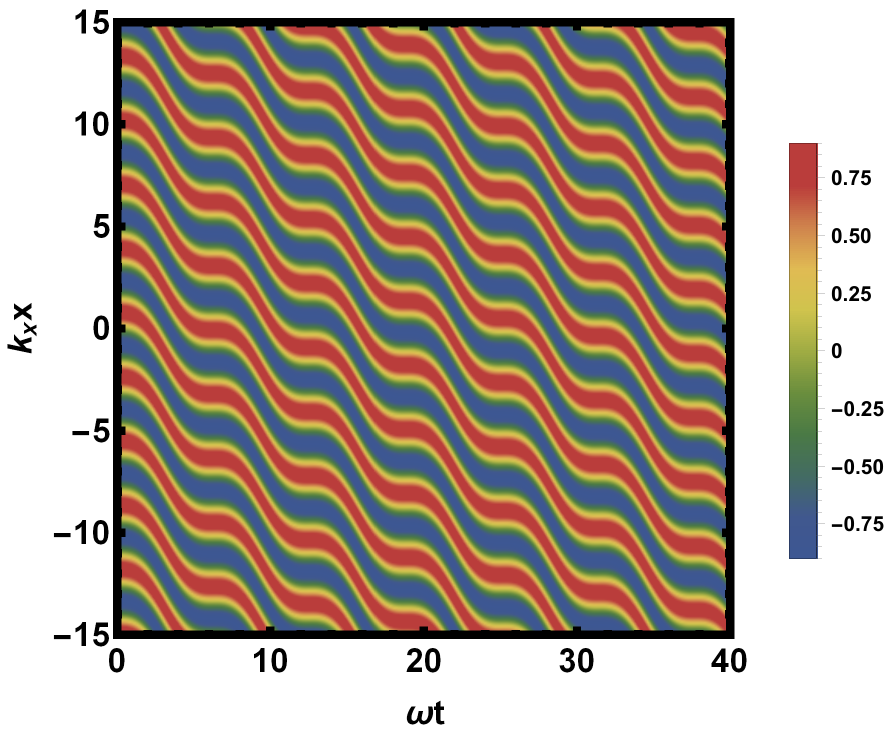}
		\label{fiht}}
	\subfloat[: $\delta L=0.8$]
{\includegraphics[width=0.5\linewidth, height=0.075\textheight]{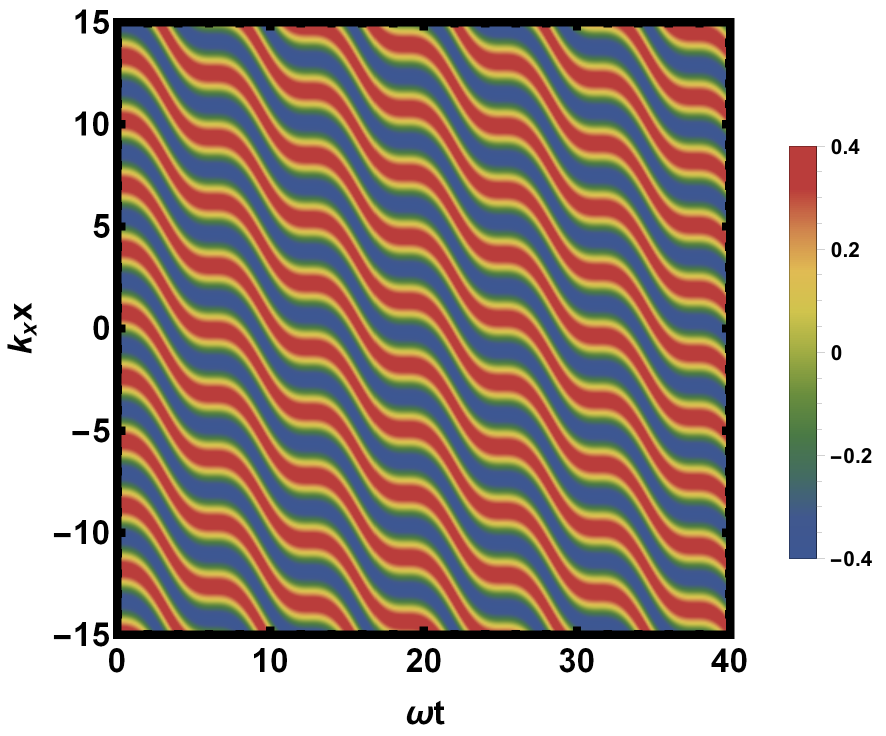}
		\label{fig1}}
	\caption{{\color{black}Current density $j_{0y}/2v_{F}$ versus time  $\omega{t}$ and wave vector ${k_{x}}x$ for $\tau_{z}=1$,  $\varepsilon L=0.9$, $\alpha=\sqrt{{L}/{2}}$, $L=1$ $\text{nm}$.   {(a)/(b)}:   ${U_{0}}v_{F}/L\hbar{\omega^{2}}=1/2$, 
			{(c)/(d)}: ${U_{0}}v_{F}/L\hbar{\omega^{2}}=2/\pi$.}}
	\label{figH}
\end{figure}

{\color{black}Fig. \ref{figH} illustrates the numerical evolution of the current density $j_{0y}/2v_F$ as a function of the time $\omega t$ and the wave vector $k_xx$, for $\tau_z = 1$, $\varepsilon L = 0.9$, $\alpha = L/2$, and $L = 1~\mathrm{nm}$. In  Figs. \ref{figH}(a,b), for $\delta L = 0$, we observe that the current density exhibits a well-defined periodic modulation along the $\omega t$ and $k_xx$ axes. The oscillatory fringes are regular and quasi-sinusoidal, indicating a regime dominated by coherent transitions induced by the electromagnetic field \cite{25}. When $\delta L = 0.8$ (Figs. \ref{figH}(c,d)), a noticeable modification of the interference pattern is observed. The oscillations become slightly distorted, and the current amplitude is redistributed, reflecting the influence of the
	parameter $\delta L$ on the electronic dynamics. Moreover, increasing the amplitude of \({U_0 v_F}/{L \hbar \omega^2}\) and \(\delta L\)  leads to an overall increase of the current oscillations. These results highlight the crucial role of the combined effect of the potential and the gap, and emphasize the importance of spatiotemporal modulation in controlling electronic transport within the system.}

\begin{figure}[H]	\centering 
		\subfloat[: $\delta L=0$ ]	{\includegraphics[width=0.475\linewidth, height=0.09\textheight]{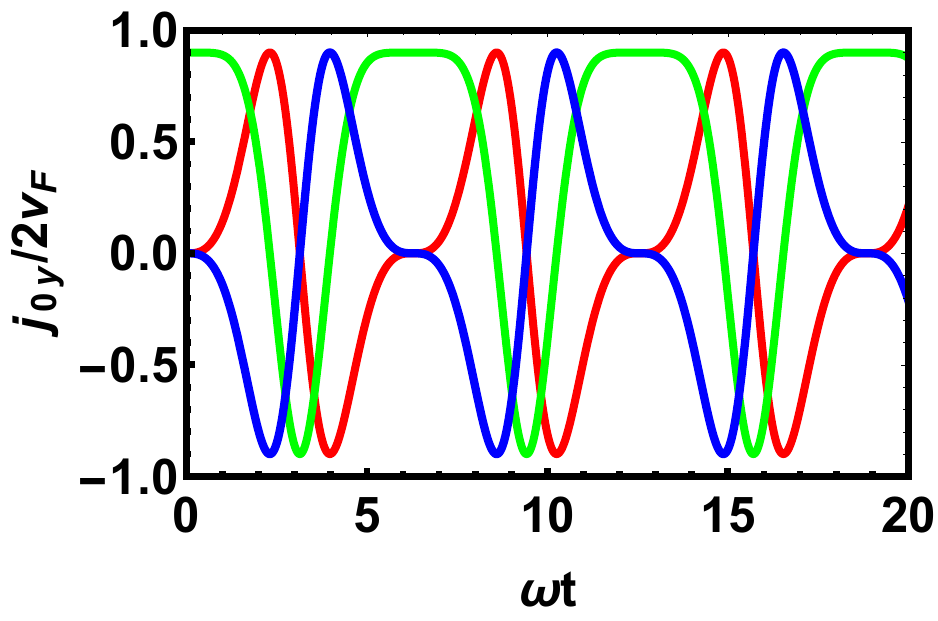}
			\label{fiz}}\subfloat[: $\delta L=0.8$]
				{\includegraphics[width=0.475\linewidth, height=0.09\textheight]{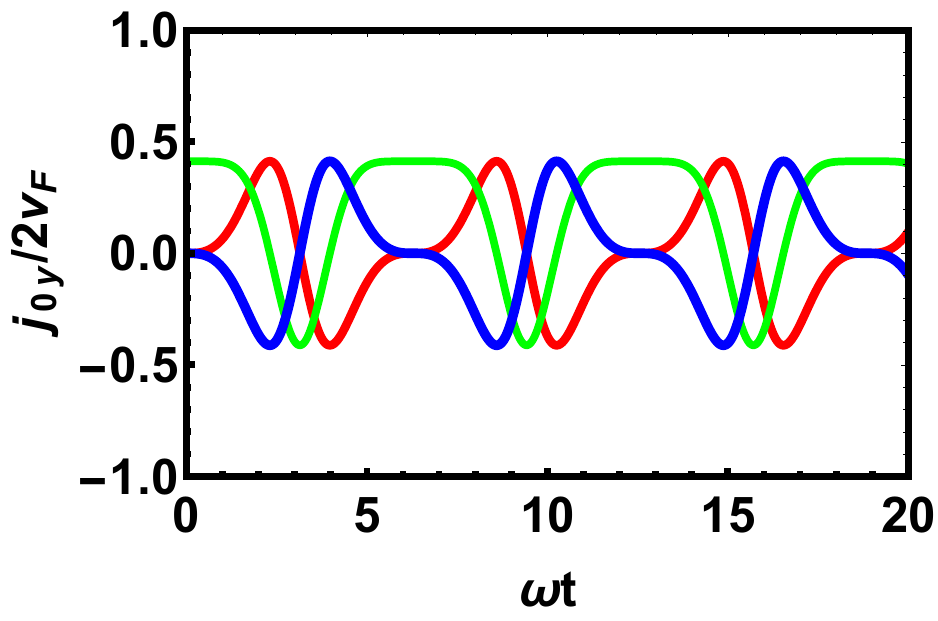}
			\label{fiz1}}
		\caption{{Current density $j_{0y}/2v_{F}$ versus time $\omega{t}$ for $\tau_{z}=1$, $\varepsilon L=0.9$, $\alpha=\sqrt{{L}/{2}}$, $L=1$ $\text{nm}$, ${U_{0}}v_{F}/L\hbar{\omega^{2}}=1/2$,   $k_{x}x=0$ (red line), $\pi/4$ (green line), $\pi/2$ (blue line).  (a): $\delta L=0$, 
				(b): $\delta L=0.8$.}}
		\label{fu1}
	\end{figure}

	To highlight the effect of the wave vector along $x$-axis on the  Josephson current, we plot in Fig.~\ref{fu1} the current density $j_{0y}/2v_{F}$ versus time. We choose   $\tau_{z}=1$,  $\varepsilon L=0.9$, $\alpha=\sqrt{{L}/{2}}$, $L=1$ $\text{nm}$ and three values of the wave vector ${k_{x}}x=0$ (red line), $\pi/4$ (green line), $\pi/2$ (blue line), with (a): $\delta L=0$, (b): $\delta L=0.6$. It is clearly seen that for ${U_{0}}v_{F}/L\hbar{\omega^{2}}=1/2$, the  periodic currents of different amplitudes oscillate between positive and negative regimes. In addition, we observe that when $\omega{t}=0$ and $\delta L=0$, $j_{0y}/2v_{F}$ converges to some values such as $j_{0y}/2v_{F}=0$ for ${k_{x}}x=0, \pi/2$ and $j_{0y}/2v_{F}=0.9$ for ${k_{x}}x=\pi/4$. However, when ${k_{x}}x=0$ and $\omega{t}=2.2$, $j_{0y}/2v_{F}$ shows a maximum, but it decreases until a minimum value at ${k_{x}}x=\pi/4, \pi/2$, and after that it also shows a continuous oscillatory behavior in both Figs. \ref{fu1}(a,b). Note that for $\delta L<\varepsilon L=0.9$, the amplitude of oscillations decreases rapidly and conversely for $\delta L>\varepsilon L$.
	   
\begin{figure}[H]\centering
		\subfloat[: ${k_{x}}x=\pi/8$]	
		{\includegraphics[width=0.475\linewidth, height=0.09\textheight]{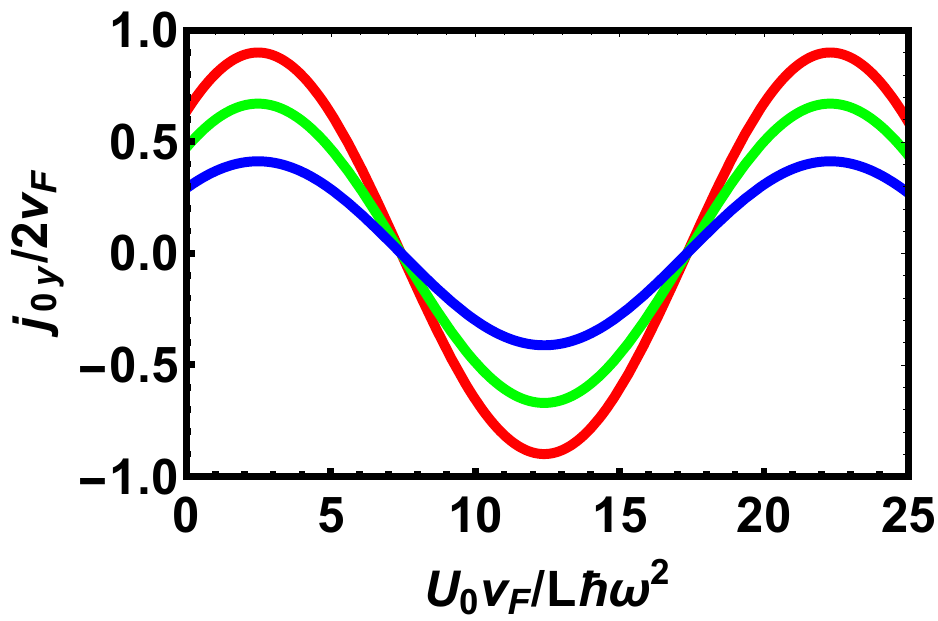}
			\label{figure5}} 
		\subfloat[: {\color{black}$\delta L=0,0.8$}]	{\includegraphics[width=0.475\linewidth, height=0.09\textheight]{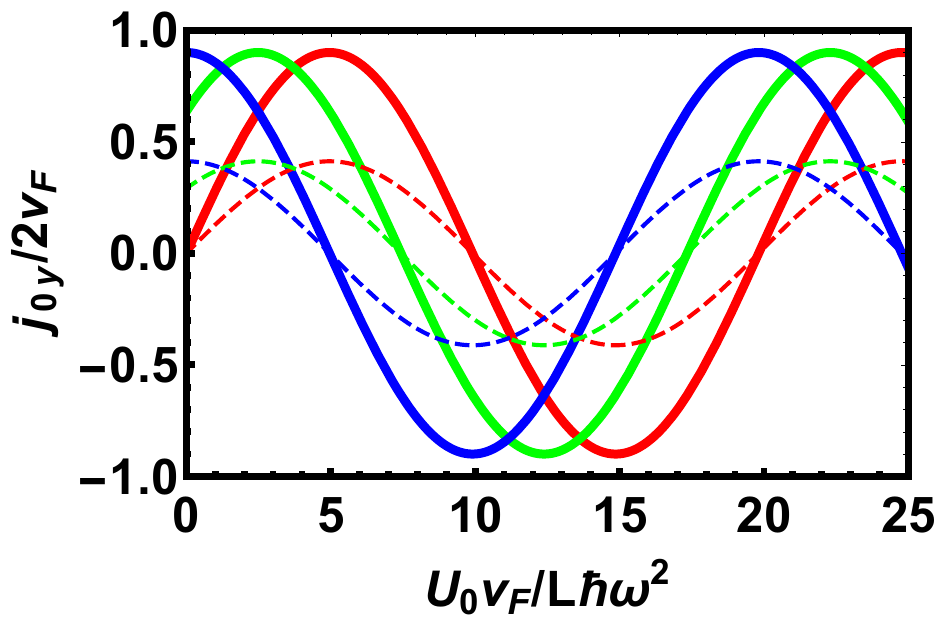}
			\label{fii}}
		\caption{{Current density $j_{0y}/2v_{F}$ versus the potential amplitude ${U_{0}}v_{F}/L\hbar{\omega^{2}}$ for $\tau_{z}=1$, $\varepsilon L=0.9$, ${\omega t}=1$, $\alpha=\sqrt{{L}/{2}}$, $L=1$ $\text{nm}$.  (a): ${k_{x}}x=\pi/8$, $\delta L= 0$ (red line), $0.6$ (green line), $0.8$ (blue line).      (b):  $\delta L=0$ (thick lines), $0.8$ (dashed lines), $k_{x}x=0$ (red line), $\pi/8$ (green line), $\pi/4$ (blue line).}}
		\label{fig2}
	\end{figure}

		In Fig.~\ref{fig2}, we present  the current density $j_{0y}/2v_{F}$ versus the potential amplitude ${U_{0}}v_{F} /L\hbar{\omega^{2}}$ for $\tau_{z}=1$, ${\omega t}=1$, $\varepsilon L=0.9$, $\alpha=\sqrt{{L}/{2}}$, $L=1$ $\text{nm}$. Fig.
	\text{\ref{fig2}}(a) {corresponds} to ${k_{x}}x=\pi/8$ and three values of the gap $\delta L= 0$ (red line), $0.6$ (green line), $0.8$ (blue line). 
	As one can see, the periodic current oscillates sinusoidally and decreases rapidly as the gap increases. In Fig.~\ref{fig2}(b), we fix two values of the parameter 
	\( \delta L \), namely \( 0 \) (thick lines) and \( 0.8 \) (dashed lines), 
	and choose \( k_{x}x = 0 \) (red line), \( \pi/8 \) (green line), and \( \pi/4 \) (blue line). 
	We observe that the periodic oscillations, which have the same amplitude, 
	start from different positive values at 
	\( {U_{0}}v_{F}/L\hbar{\omega^{2}} = 0 \) and become shifted as \( k_{x}x \) increases. 
	However, for larger values of \( {U_{0}}v_{F}/L\hbar{\omega^{2}} \) 
	or when \( \delta L \geq 0 \), the oscillation amplitude decreases. 
	{Consequently, the Josephson-like oscillatory behavior becomes less pronounced, 
		indicating a suppression of coherent phase-like dynamics in the system.}
	
\begin{figure}[H]\centering
	\subfloat[: $\delta L=0$]
	{\includegraphics[width=0.475\linewidth, height=0.09\textheight]{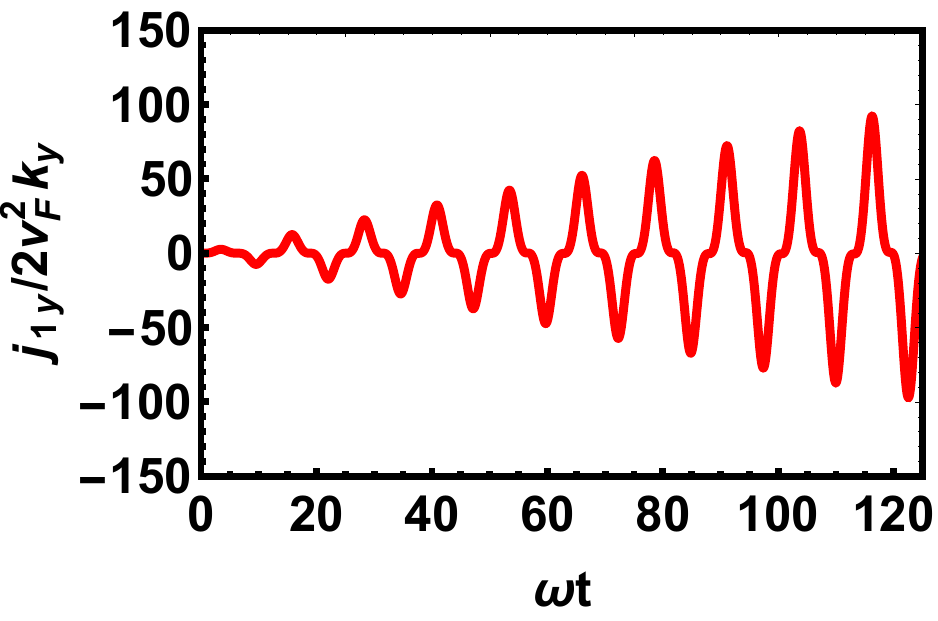}
		\label{fdi23}} 
	\subfloat[: $\delta L=0.6$]{\includegraphics[width=0.475\linewidth, height=0.09\textheight]{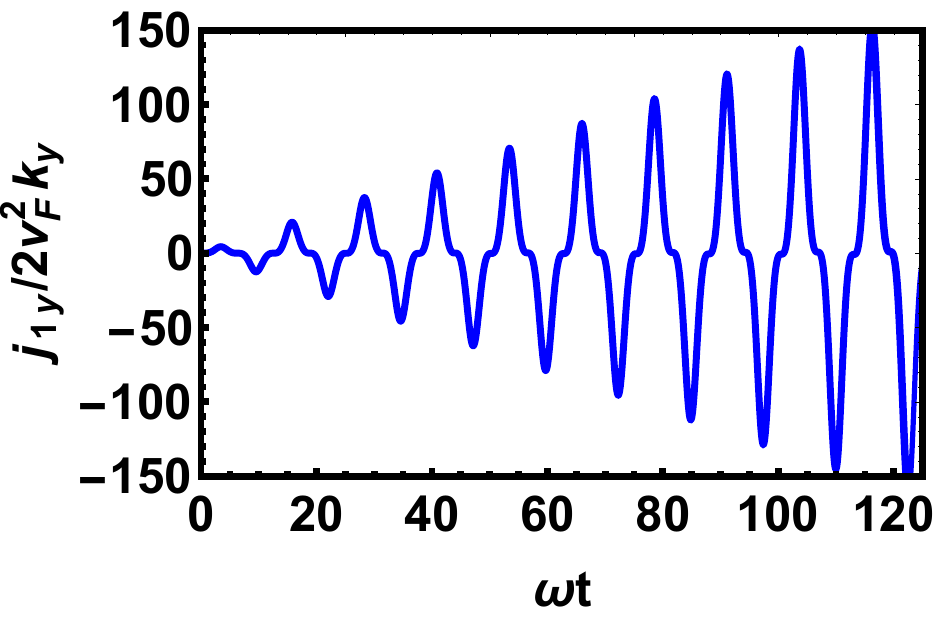}
		\label{fdi25}}\\
	\subfloat[: $\delta L=0$]{\includegraphics[width=0.475\linewidth, height=0.09\textheight]{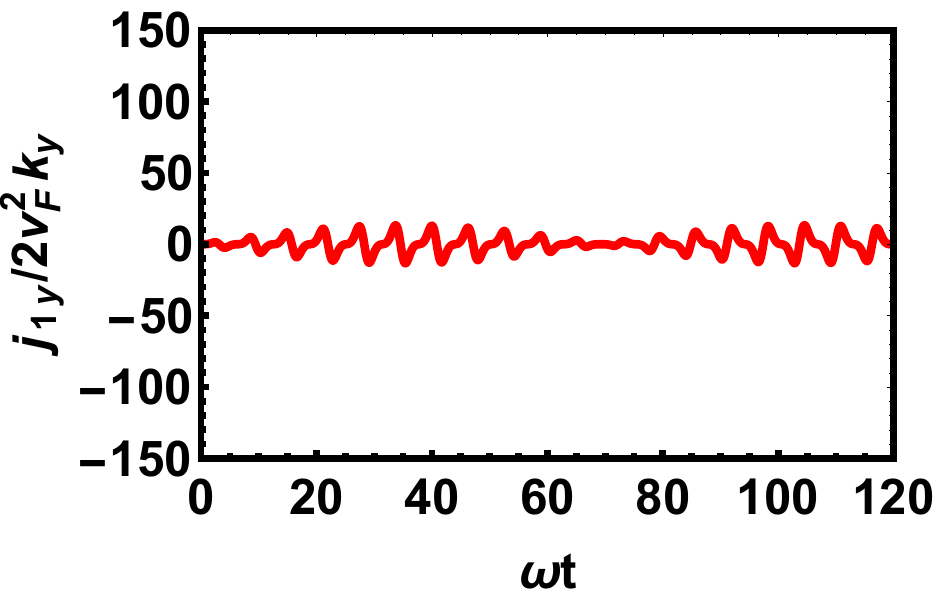}
		\label{fdi2}}
	\subfloat[: $\delta L=1.2$]{\includegraphics[width=0.475\linewidth, height=0.09\textheight]{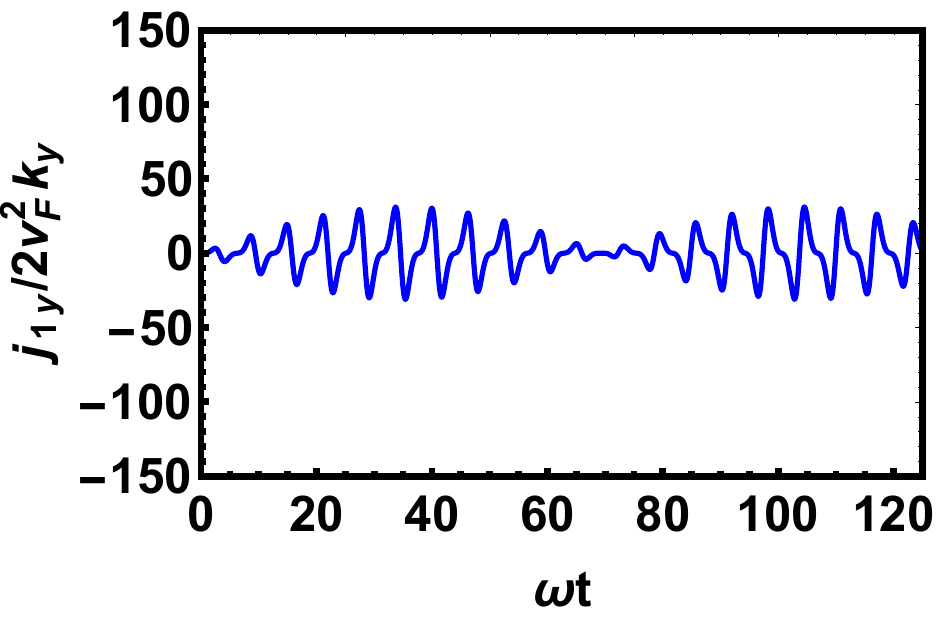}
		\label{fdi2r}}
	\caption{{Current density $j_{1y}/2v^{2}_{F}k_{y}$ versus time  $\omega{t}$ for $\varepsilon L=0.9$, $x=0$, $k_{x}=0$, $\alpha=\sqrt{{L}/{2}}$, $L=1$ $\text{nm}$, $\omega=1$ \text{$s^{-1}$}.  (a,b):  ${U_{0}}v_{F}/L\hbar{\omega^{2}}=1/2$, $n=0.9999$. $\delta L=0$ (red line), $0.6$ (blue line). (c,d):  ${U_{0}}v_{F}/L\hbar{\omega^{2}}=3/\pi$, $n=2$, $\delta L= 0$ (red line), $1.2$ (blue line).}}
	\label{hde4}
\end{figure}

Fig.~\ref{hde4} presents the time evolution of the current density $j_{1y}/2v_{F}^{2}k_{y}$, computed for the parameter set $\varepsilon L = 0.9$, $\alpha = \sqrt{L/2}$, $L = 1~\mathrm{nm}$, $\omega = 1~\mathrm{s}^{-1}$, $x = 0$, and $k_{x} = 0$. We will adjust the value of the energy gap \( \delta L \) in further analysis.  
Under Shapiro resonance conditions as shown in Figs.~\ref{hde4}(a,b) with \( U_{0}v_{F} /L\hbar\omega^{2} = 1/2 \), the current maintains uniform oscillations that switch between positive and negative. Importantly, these oscillations exhibit increasing amplitude over time, which aligns with the resonant behavior noted in  (\ref{k1ju}).  
Figs.~\ref{hde4}(c,d) illustrate the current response at off-resonance conditions for \( U_{0}v_{F} /L\hbar\omega^{2} = 3/\pi \). Current response appears altered, as oscillations exhibit rapid initial growth before gradual decay toward near-zero over the interval \( 60 \leq \omega t \leq 80 \). Further temporal advancement reveals the emergence of new oscillations, but these are now reverse to the prior forward progression observed for \( \omega t \leq 73 \).
For the case \( \delta L = 0 \), the current behavior matches well with that described in  \cite{25}, although the rate of increase of the amplitude is sharper.



\begin{figure}[H]\centering
		\subfloat[: $\delta L=0$]{\includegraphics[width=0.475\linewidth, height=0.09\textheight]{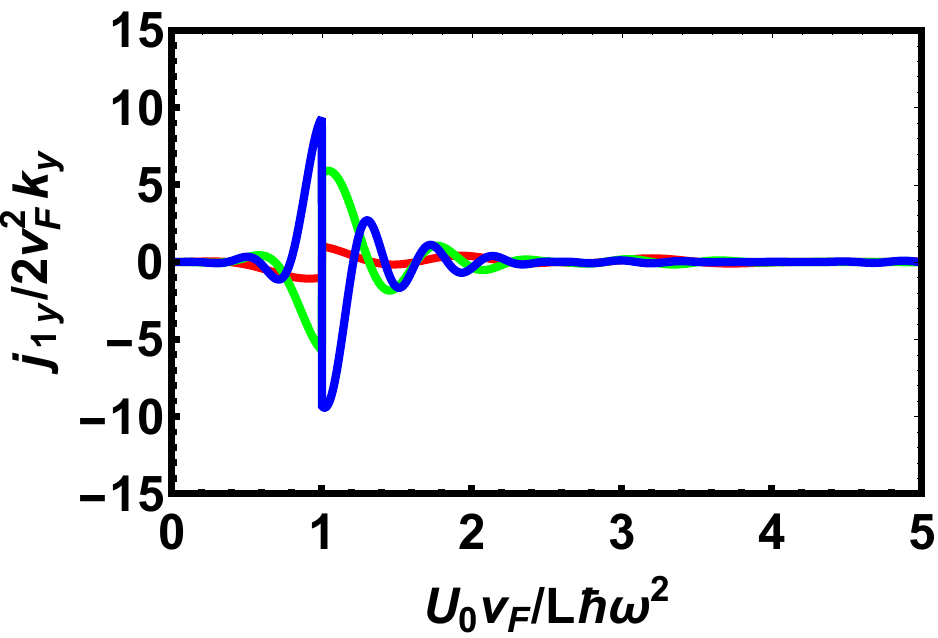}
			\label{figur5}}
		\subfloat[: $\delta L=0.5$]{\includegraphics[width=0.475\linewidth, height=0.09\textheight]{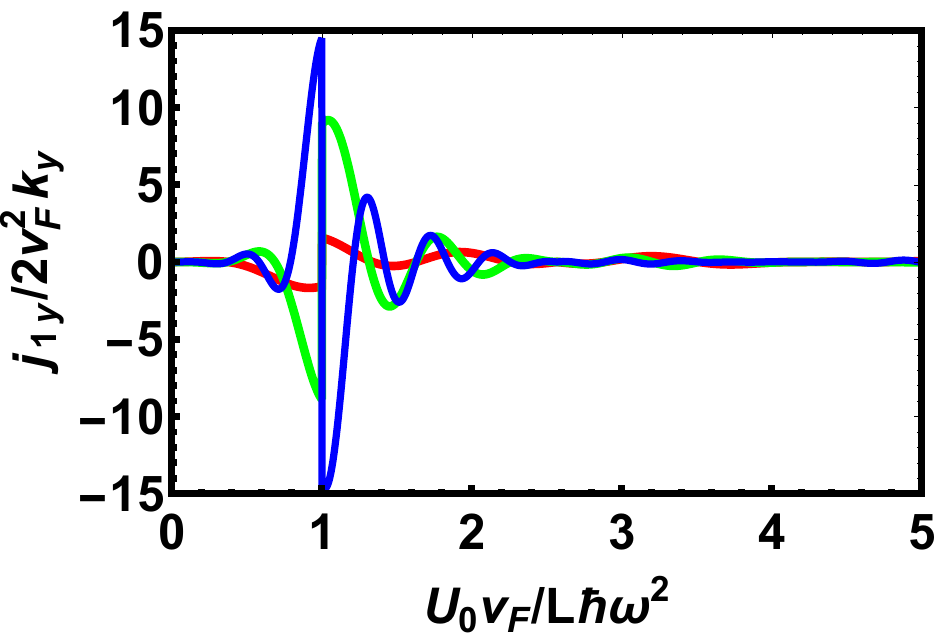}
			\label{fii5}}
		\caption{Current density $j_{1y}/2v^{2}_{F}k_{y}$ versus the potential amplitude ${U_{0}}v_{F}/L\hbar{\omega^{2}}$ for $\varepsilon L=0.9$, $x=0$, $k_{x}=0$, $\alpha=\sqrt{{L}/{2}}$, $L=1$ $\text{nm}$, $\omega=1$ \text{$s^{-1}$}, $n=2$, ${\omega t}=5$ (red line), $10$ (green line), $15$ (blue line).  (a){:} $\delta L=0$,  (b){:} $\delta L=0.5$.}
		\label{figh2}
	\end{figure}

Fig. \ref{figh2} displays the current density $j_{1y}/2v^{2}_{F}k_{y}$ versus the potential amplitude ${U_{0}}v_{F}/L\hbar{\omega^{2}}$ for $\varepsilon L=0.9$, $x=0$, $k_{x}=0$, $\alpha=\sqrt{{L}/{2}}$, $L=1$ $\text{nm}$, $\omega=1$ \text{$s^{-1}$}, $n=2$ and three values of time ${\omega t}=5$ (red line), $10$ (green line), $15$ (blue line), with (a){:} $\delta L= 0$ and(b){:} $\delta L= 0.5$. 
%
In both cases, the current density exhibits oscillations between positive and negative values, with amplitudes increasing noticeably as time progresses. However, for \( \frac{U_{0}v_{F}}{L\hbar\omega^{2}} > 2 \), the amplitude of the oscillations decreases significantly and eventually approaches zero.
	Furthermore, increasing the Shapiro step index \( n > 2 \) results in a rightward shift of the oscillatory pattern, indicating that higher steps require larger potential amplitudes to maintain resonance. Comparing Figs.~\ref{figh2}(a) and \ref{figh2}(b), we observe that introducing a finite gap (\( \delta L = 0.5 \)) leads to sharper and more pronounced peaks in the current response. In both cases, the overall magnitude of \( j_{1y}/2v_{F}^{2}k_{y} \) continues to grow with time, emphasizing the cumulative effect in the Josephson-like response

\begin{figure}[H]\centering
			\subfloat[: ${U_{0}}v_{F}/L\hbar{\omega^{2}}=1/\pi$]{\includegraphics[width=0.475\linewidth, height=0.09\textheight]{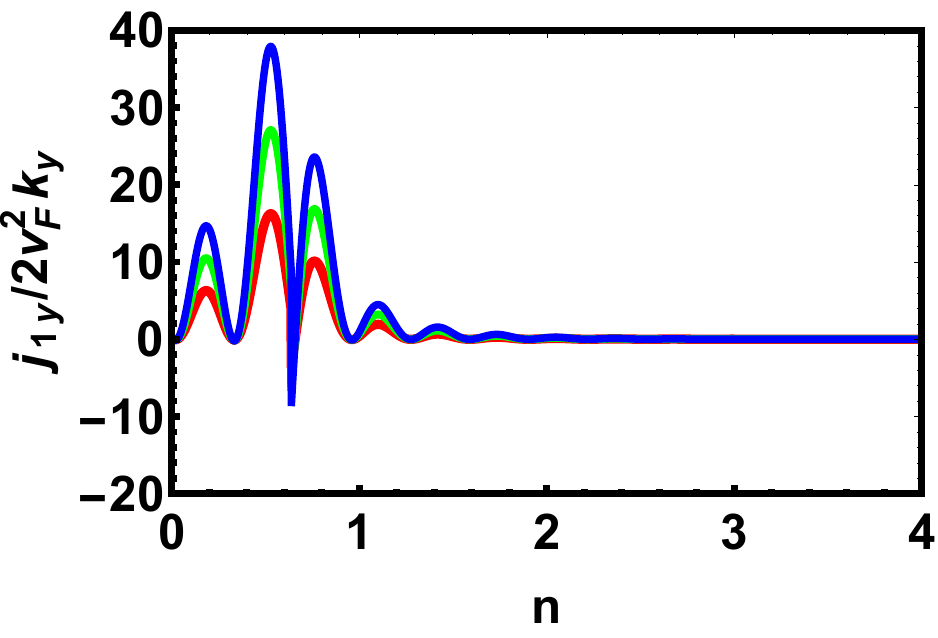}
				\label{figur}}
			\subfloat[: ${U_{0}}v_{F}/L\hbar{\omega^{2}}=3/\pi$]{\includegraphics[width=0.475\linewidth, height=0.09\textheight]{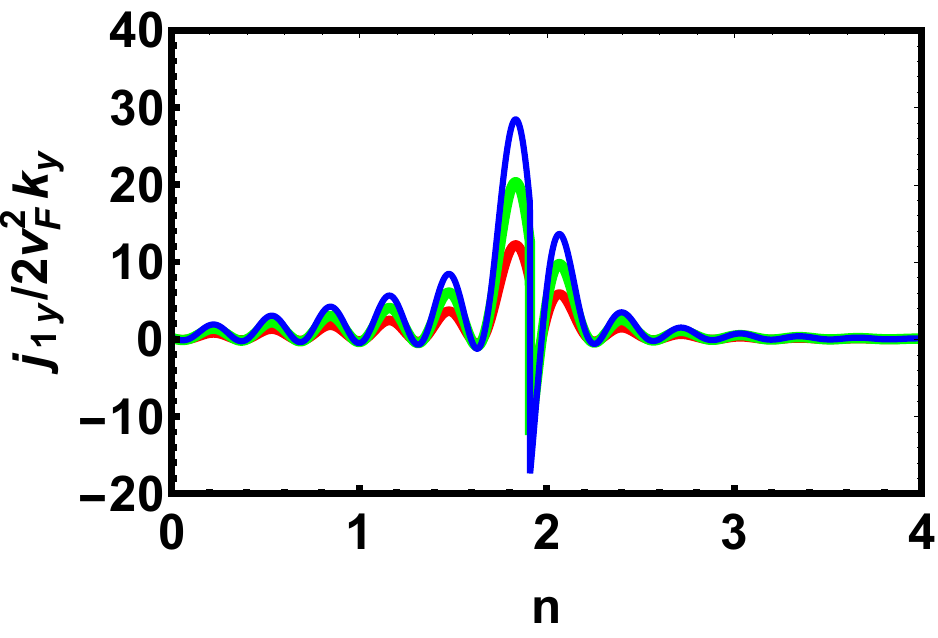}
				\label{fi5}}
			\caption{Current density $j_{1y}/2v^{2}_{F}k_{y}$ versus Shapiro steps $n$ for $\varepsilon L=0.9$, $x=0$, $k_{x}=0$, $\alpha=\sqrt{{L}/{2}}$, $L=1$ $\text{nm}$, $\omega=1$ \text{$s^{-1}$}, ${\omega t}=20$, $\delta L=0$ (red line), $0.6$ (green line), $1.2$ (blue line).  (a){:} ${U_{0}}v_{F}/L\hbar{\omega^{2}}=1/\pi$,  (b){:} ${U_{0}}v_{F}/L\hbar{\omega^{2}}=3/\pi$.}
			\label{figh}
		\end{figure}

	Fig.~\ref{figh} shows how the current density \( j_{1y}/2v^{2}_{F}k_{y} \) changes as a function of the Shapiro step index \( n \), outside the resonance condition (\ref{k1ju}). The parameters of the system are chosen as follows: \( \varepsilon L = 0.9 \), \( x = 0 \), \( k_{x} = 0 \), \( \alpha = \sqrt{L/2} \), \( L = 1~\text{nm} \), \( \omega = 1~\text{s}^{-1} \), and \( \omega t = 20 \). Three values of the energy gap are considered: \( \delta L = 0 \) (red curve), \( 0.6 \) (green curve), and \( 1.2 \) (blue curve).
	In Fig.~\ref{figh}(a), for the potential amplitude \( U_{0}v_{F}/L\hbar\omega^{2} = 1/\pi \), we see that the current density grows with higher values of the Shapiro step \( n \) and reaches the maximum value around \( j_{1y}/2v^{2}_{F}k_{y} = 39 \) for \( \delta L = 1.2 \). For small \( n < 1 \), in contrast, the current density is weak and limited to the negative side. After the maximum, it falls off gradually and eventually disappears.
	Fig.~\ref{figh}(b) presents the situation for a larger potential amplitude \( U_{0}v_{F}/L\hbar\omega^{2} = 3/\pi \). In this case, the current density also grows with \( n \), albeit more slowly than before. It rises to a maximum value of approximately \( j_{1y}/2v^{2}_{F}k_{y} = 29 \) before falling to close to zero for \( n \ge 3 \). We also see a significant increase in the number and complexity of oscillations over Fig.~\ref{figh}(a), reflecting a more complex response for the larger driving amplitude.
	In general, these findings indicate that the Josephson-like current is very sensitive to the Shapiro step and the energy gap. Specifically, the existence of a finite gap \( (0 < \delta L < 1) \) considerably increases the current density prior to ultimate suppression at large \( n \).
	
	\section{Spatially-temporally periodic potential}\label{IV}
	 
\begin{widetext} 
		We consider the second example of potential  
	$U(x,t)=U_{0}\cos\frac{x}{L}\cos{\omega t}$ periodic in space with $2\pi L$ and in time with $2\pi/\omega$. After calculation, we end up with the current density
		\begin{align}\label{hy17}
			j_{0y}=4v_{F}\tau_{z}\alpha^{2}
			\sqrt{\left|{\varepsilon^{2}-\delta^{2}}\right|}
			\sin\left[2k_{x}x-\frac{4U_{0}Lv_{F}\sin\left(\frac{x}{L}\right)}{\hbar\omega^{2}L^{2}-\hbar v_{F}^{2}}\sin\left[\frac{\omega L+v_{F}}{2L}t\right]\sin\left[\frac{\omega L-v_{F}}{2L}t\right]\right].
		\end{align}
	\end{widetext}

\begin{figure}[H]\centering
	\subfloat[: $k_{x}x=0$]{\includegraphics[width=0.475\linewidth, height=0.09\textheight]{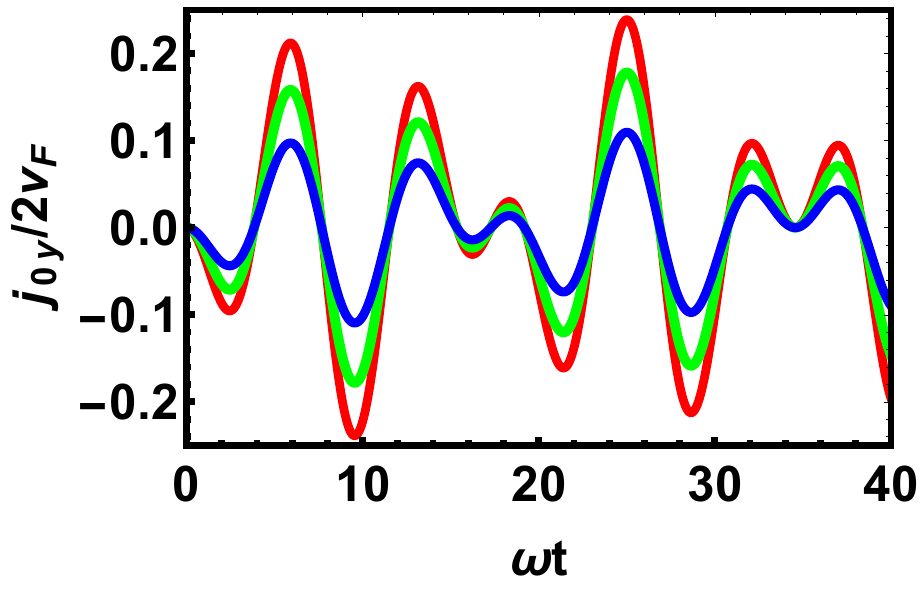}
		\label{figure11}}
	\subfloat[: {\color{black}$k_{x}x=-\pi/8$}]{\includegraphics[width=0.475\linewidth, height=0.09\textheight]{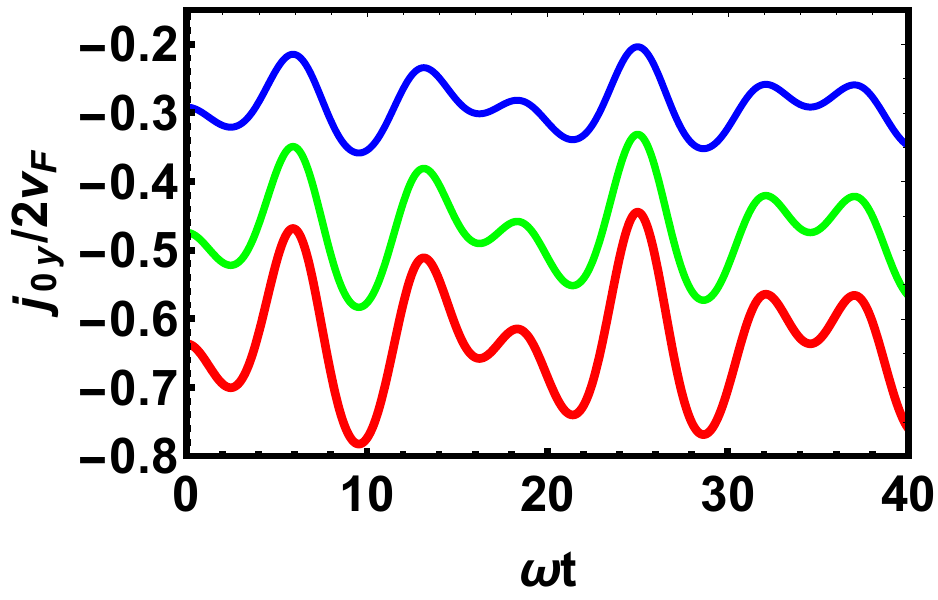}
		\label{figure40}}\\
	\subfloat[: $k_{x}x=0$]{\includegraphics[width=0.475\linewidth, height=0.09\textheight]{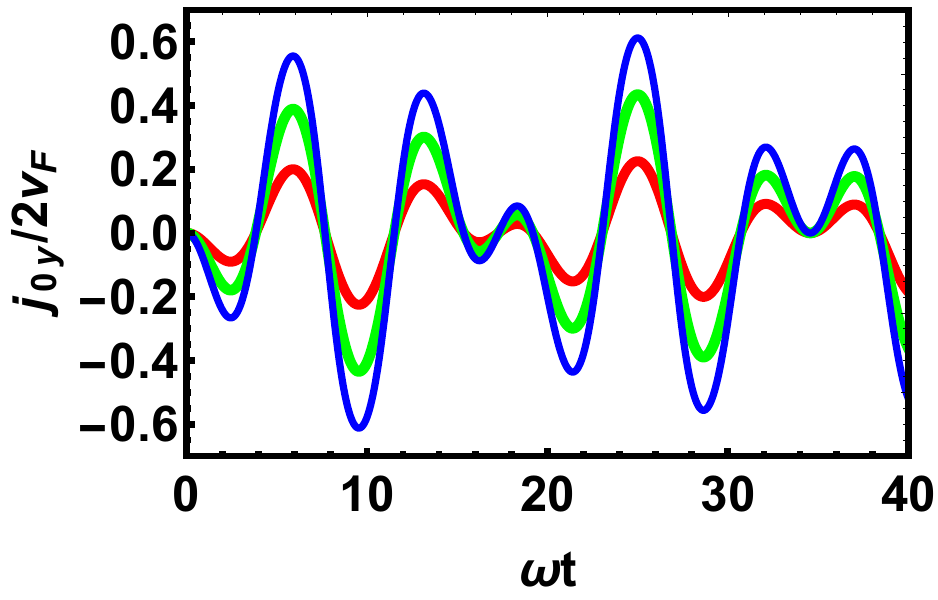}
		\label{figu1}}
	\subfloat[: {\color{black}$k_{x}x=-\pi/8$}]{\includegraphics[width=0.475\linewidth, height=0.09\textheight]{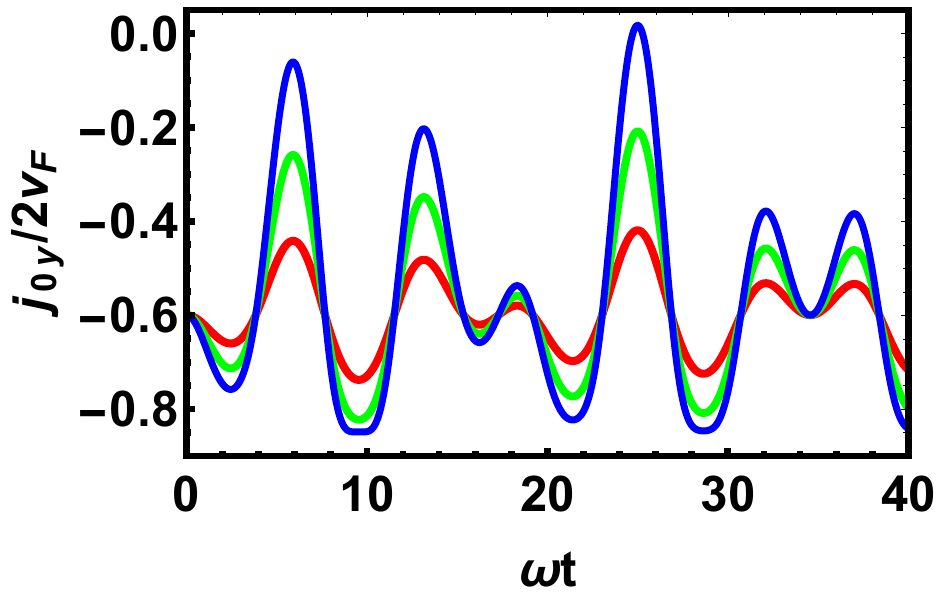}
		\label{figu}}
	\caption{{Current density $j_{0y}/2v_{F}$ versus time $\omega{t}$ for $\tau_{z}=1$, $x= \pi L/2$, $\omega=\pi v_{F} /2L$, $\varepsilon L=0.9$, $\alpha=\sqrt{{L}/{2}}$, $L=1$ $\text{nm}$. (a,b){:} $U_{0}L/\hbar v_{F}=0.1$, $\delta L=0$ (red line), $0.6$ (green line), $0.8$ (blue line). (c,d){:} $\delta L=0.3$, $U_{0}L/\hbar v_{F}=0.1$ (red line), $U_{0}L/\hbar v_{F}=0.2$ (green line),  $U_{0}L/\hbar v_{F}=0.3$ (blue line).}}
	\label{figure}
\end{figure}
	
To illustrate the effect of the gap and potential amplitude on the Josephson current,  in Fig. \text{\ref{figure}}, we present the current density $j_{0y}/2v_{F}$ (\ref{hy17}) versus time when $\varepsilon L>\delta L$ and away from resonance $\omega=\pi v_{F}/2L$ for $\tau_{z}=1$, $x=\pi L/2$, $\varepsilon L=0.9$, $\alpha=\sqrt{{L}/{2}}$, $L=1$ $\text{nm}$. Indeed, in Figs. \text{\ref{figure}}(a,b) {we choose $U_{0}L/\hbar v_{F}=0.1$} and three values of the gap $\delta L=0$ (red line), $0.6$ (green line), $0.8$ (blue line). For $\delta L=0$, we reproduce the results obtained in \cite{25}. For $\delta L\neq0$, we observe that the aperiodic oscillations increase negatively when $k_{x}x=0$ and $0\leq \omega{t} \leq 4$. After these values, they are oscillating from positive to negative regimes. For {\color{black}$k_{x}x=-\pi/8$}, when $\omega t$ increases the current density starts from different points with {\color{black}negative values}. From Figs. \text{\ref{figure}}(a,b) {we} notice that when $\delta L$ increases, {\color{black}$j_{0y}/2v_{F}$ decreases  at $k_xx = 0$, while it increases at $k_xx = -\pi/8$.}  Figs.  \text{\ref{figure}}(c,d) {correspond} to $\delta L=0.3$ with three values of the potential amplitude $U_{0}L/\hbar v_{F}=0.1$ (red line), $0.2$ (green line), $0.3$ (blue line). It is clearly see that when $U_{0}L/\hbar v_{F}$ and $\omega{t}$ increase $j_{0y}/2v_{F}$ also increases quickly in negative and positive regimes. Additionally, one sees that the starting point of these oscillations is $j_{0y}/2v_{F}=0$ for $k_{x}x=0$ and {\color{black}$j_{0y}/2v_{F}=-0.6$} for {\color{black}$k_{x}x=-\pi/8$}, this increase depends on the value of the wave vector along $x$-axis.  
	
	\begin{figure}[H]\centering
		\subfloat[: $\delta L=0$]{\includegraphics[width=0.475\linewidth, height=0.079\textheight]{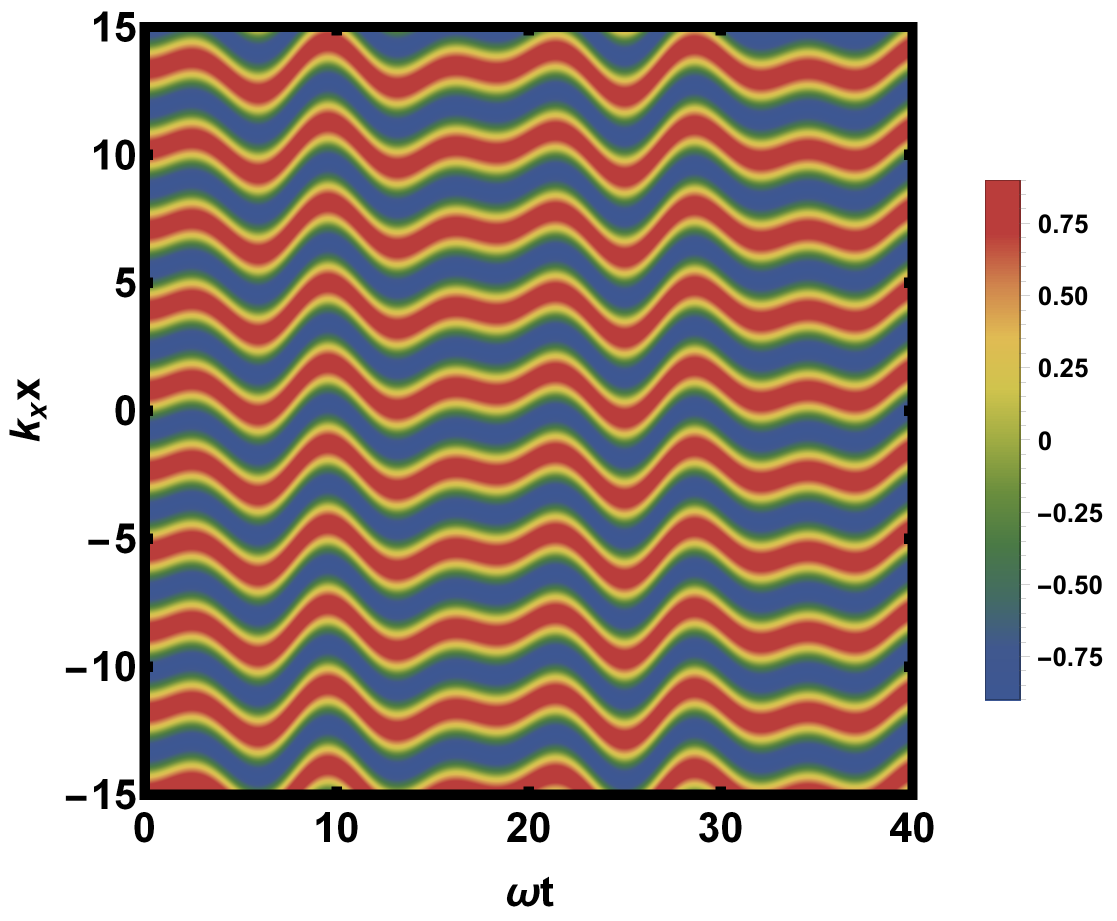}
			\label{figur11}}
		\subfloat[: $\delta L=0.8$]{\includegraphics[width=0.475\linewidth, height=0.079\textheight]{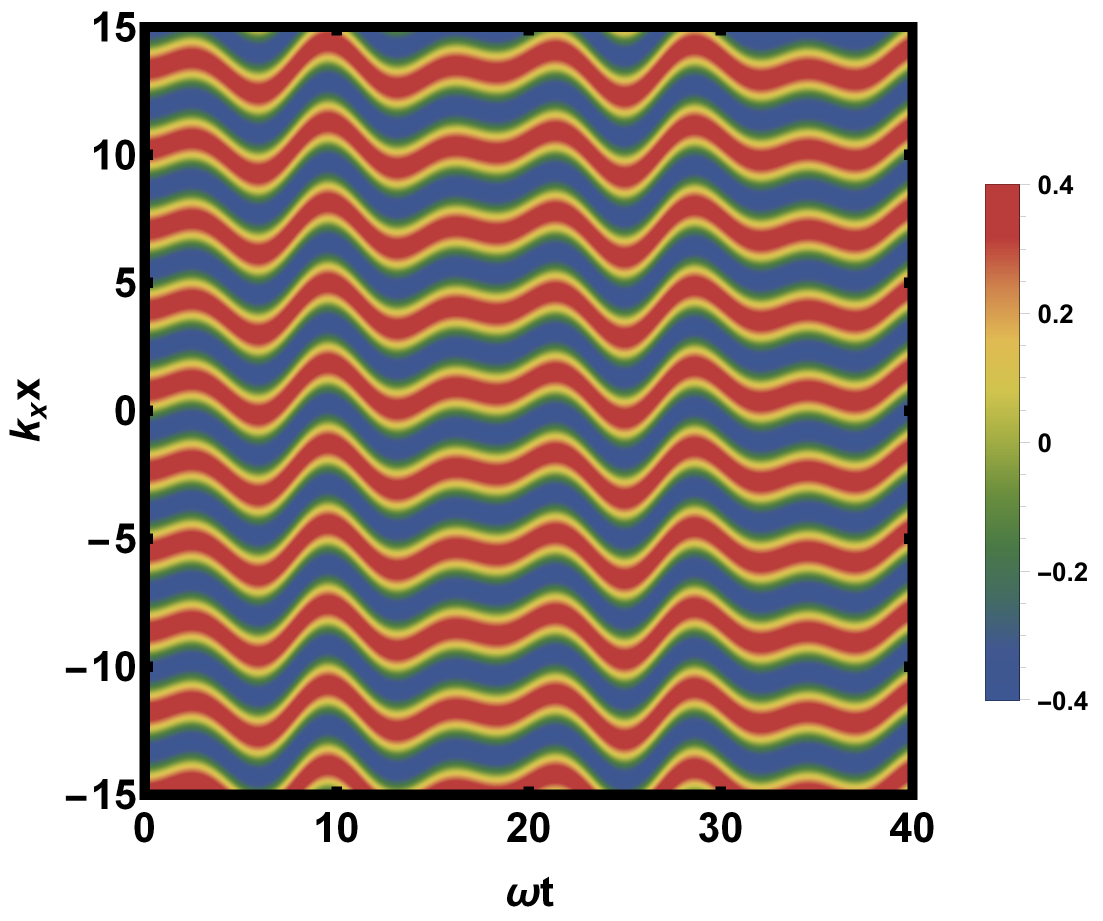}
			\label{figre40}}\\
		\subfloat[: $\delta L=0$]{\includegraphics[width=0.475\linewidth, height=0.079\textheight]{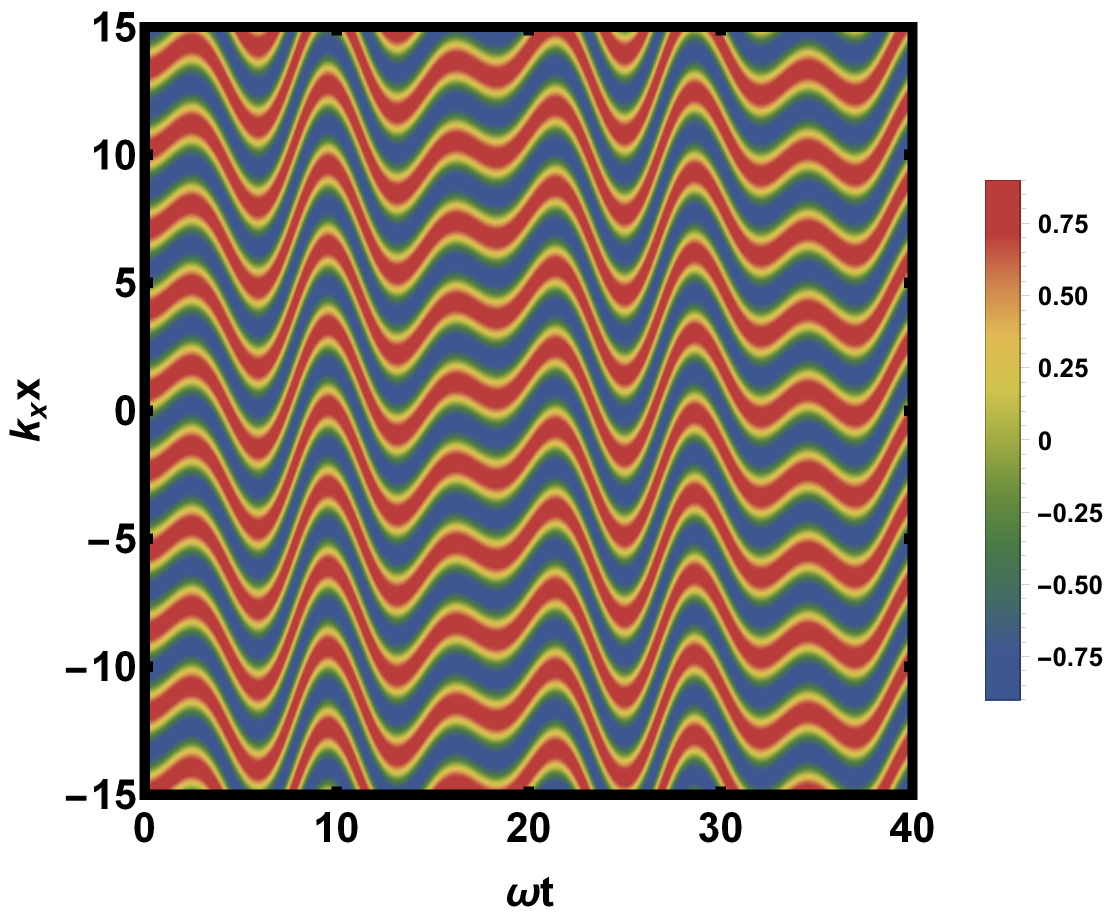}
			\label{figt}}
		\subfloat[: $\delta L=0.8$]{\includegraphics[width=0.475\linewidth, height=0.079\textheight]{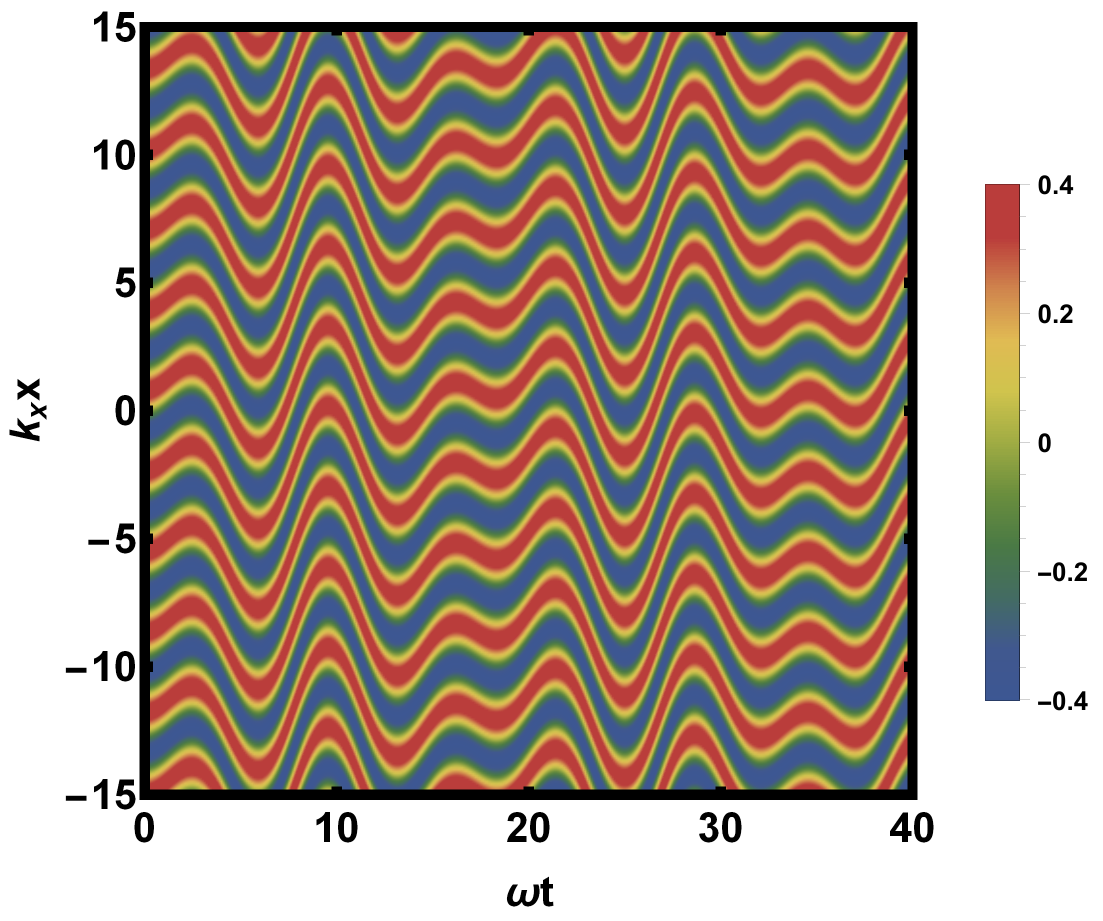}
			\label{figr1}}
		\caption{{\color{black}Current density $j_{0y}/2v_{F}$ versus time $\omega{t}$ and wave vector $k_xx$ for $\tau_{z}=1$, $x=\pi L/2$, $\omega=\pi v_{F} /2L$, $\varepsilon L=0.9$, $\alpha=\sqrt{{L}/{2}}$, $L=1$ $\text{nm}$. 
				(a)/(b){:} $U_{0}L/\hbar v_{F}=0.8$.   (c)/(d){:}  $U_{0}L/\hbar v_{F}=1.9$.}}
		\label{fig}
	\end{figure}
	To better understand how the wave vector, energy gap, and potential amplitude affect the Josephson-like current, we plot the normalized current density $j_{0y}/2v_{F}$ as a function of time in Fig.~\ref{fig}. The system parameters are set to $\tau_{z} = 1$, $x = \pi L/2$, $\varepsilon L = 0.9$, {\color{black}$\delta L =0, 0.8$}, $\alpha = \sqrt{L/2}$, and $L = 1~\mathrm{nm}$, with the driving frequency intentionally chosen away from resonance, specifically $\omega = \pi v_{F}/2L$.
	In Figs.~\ref{fig}(a,b), we examine the current response under a moderate potential amplitude {\color{black}$U_{0}L/\hbar v_{F} = 0.8$}. 
	{\color{black}The current density exhibits a clear oscillatory behavior in both time and wave-vector domains \cite{25}. As shown, increasing the wave vector leads to more pronounced aperiodic oscillations and} sharp peaks begin to appear superimposed on the oscillatory background, a feature not observed at lower values. 
	{\color{black}This behavior originates from the phase accumulation of the electronic wave functions and the interference between propagating states at the fixed position $x=\pi L/2$. The spatial oscillations demonstrate that the magnitude and sign of the current are highly sensitive to the longitudinal wave vector.}
	Figs.~\ref{fig}(c,d) examine how the evolution of the system changes when the potential amplitude is raised to {\color{black}$U_{0}L/\hbar v_{F} =1.9$}. With this stronger drive, the current density shows more intense and vigorous oscillations. A significant change is seen {\color{black}with the increase of  $k_{x}x$ values}, where sharp peaks start to develop that extend into the regime of negative current. 
	Such reversals in current direction are most apparent in the time ranges $8 \leq \omega t \leq 12$, $21 \leq \omega t \leq 24$, and $28 \leq \omega t \leq 31$, showing that the system experiences alternating flow patterns as a function of both time and driving strength. 
	Additionally, when the gap is further raised to {\color{black}$\delta L = 0.8$}, as shown in Figs.~\ref{fig}(b,d), the amplitude of the current oscillations significantly decreases over time. This decrease highlights the damping role played by the energy gap, which appears to suppress the system’s ability to sustain large current responses under periodic driving. Collectively, these findings demonstrate the delicate balance between potential strength, wave vector, and energy gap in shaping the temporal dynamics of Josephson-like currents.


	We now examine two limiting cases. The first corresponds to $\omega =0$, in which  the current density~(\ref{hy17}) simplifies to
	\begin{widetext}
	\begin{align}\label{ht}
		j_{0y}=4v_{F}\tau_{z}\alpha^{2}
		\sqrt{\left|{\varepsilon^{2}-\delta^{2}}\right|}\sin\left[2k_{x}x-{4\frac{U_{0}L}{\hbar v_{F}}\sin\left(\frac{x}{L}\right)}\sin^{2}\left(\frac{v_{F}t}{2L}\right)\right].
	\end{align} 	
	\end{widetext}
	It exhibits oscillations with a frequency \( v_{F}/L \), which depends on the Fermi velocity. 
	{This behavior reflects coherent oscillations in the system~\cite{28,29}.} 
	The second limiting case involves \( \omega \longrightarrow \pm v_{F}/L \), and then we have
	\begin{align}\label{eq26}
		j_{0y}=4v_{F}\tau_{z}\alpha^{2}
		\sqrt{\left|{\varepsilon^{2}-\delta^{2}}\right|}\sin\left[2k_{x}x\mp \frac{tU_{0}}{\hbar}\sin\left(
		\frac{x}{L}\right)\sin\omega t\right]
	\end{align} 
	which becomes null  for $\varepsilon=\delta$.   Moreover, the current density results reported in \cite{25} are recovered in the limit $\delta=0$.
	
		\begin{figure}[H]\centering
		\subfloat[: $k_{x}x=0$]{\includegraphics[width=0.475\linewidth, height=0.09\textheight]{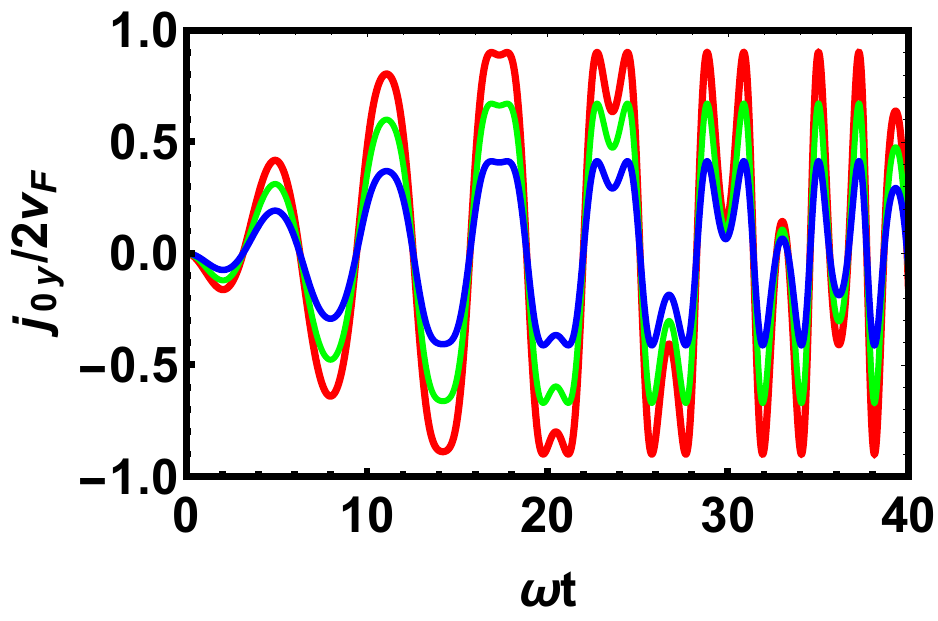}
			\label{fi23}}
		\subfloat[: {\color{black}$k_{x}x=-\pi/8$}]
		{\includegraphics[width=0.475\linewidth, height=0.09\textheight]{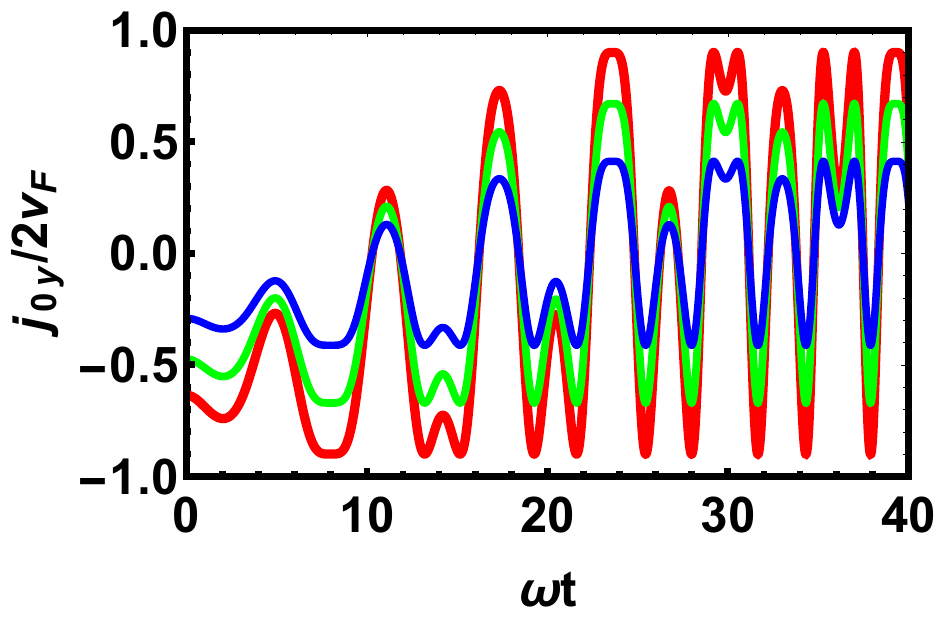}
			\label{fi24}}	
		\caption{Current density $j_{0y}/2v_{F}$ versus time $\omega{t}$ for  $\tau_{z}=1$, $x=\pi L/2$, $\omega=v_{F}/L$,  $\varepsilon L=0.9$, $U_{0}L/\hbar v_{F}=0.1$,  $\alpha=\sqrt{{L}/{2}}$, $L=1$ $\text{nm}$, $\delta L=0$ (red line), $0.6$ (green line), $0.8$ (blue line). (a){:} $k_{x}x=0$, (b){:}  {\color{black}$k_{x}x=-\pi/8$}.}
		\label{he4}
	\end{figure} 
	To show the influence of the introduced gap on the current density, we illustrate in Fig. \ref{he4} $j_{0y}/2v_{F}$ (\ref{eq26}) versus  time for $\tau_{z}=1$, $x= \pi L/2$,  $\varepsilon L=0.9$, $U_{0} L/\hbar v_{F}=0.1$, $\alpha=\sqrt{{L}/{2}}$, $L=1$ $\text{nm}$ and three values of the gap $\delta L=0$ (red line), $0.6$ (green line), $0.8$ (blue line) at  resonance condition $\omega=v_{F}/L$. Indeed, we observe that in Fig. \text{\ref{he4}}(a) {for} $\delta L=0$ and when $k_{x}x=0$, the aperiodic oscillations grow resonantly with time  for small amplitude $U_{0}L/\hbar v_{F}=0.1$ but one can see that by increasing the values of $\delta L$, the amplitude of these oscillations decreases rapidly. Whereas for {\color{black}$k_{x}x=-\pi/8$}, we notice that in Fig. \text{\ref{he4}}(b) {$j_{0y}/2v_{F}$} {\color{black}starts from negative values at $\omega{t}=0$} and exhibits an oscillatory behavior with the appearance of the sharp peaks for $\omega{t}\geq 37$ {\color{black}\cite{25}}. Therefore, we deduce that the behavior of current densities shows different apperiodicity and so the phenomenon of resonance becomes uninteresting.
	
	\begin{figure}[H]\centering
	\subfloat[: $\delta L=0$]{\includegraphics[width=0.475\linewidth, height=0.08\textheight]{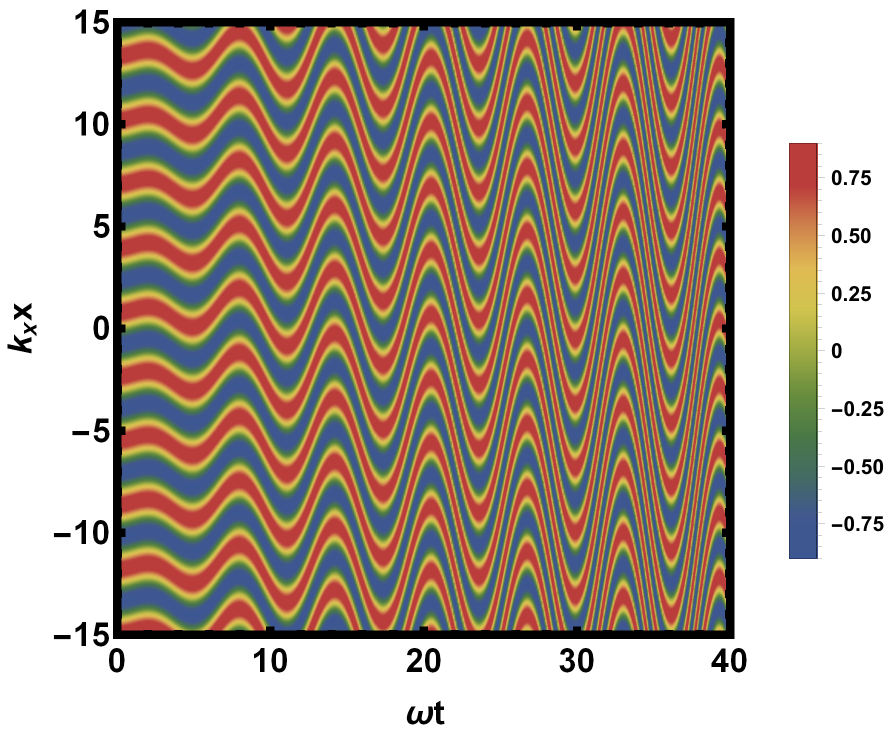}
		\label{figur1}}
	\subfloat[: $\delta L=0.8$]{\includegraphics[width=0.475\linewidth, height=0.08\textheight]{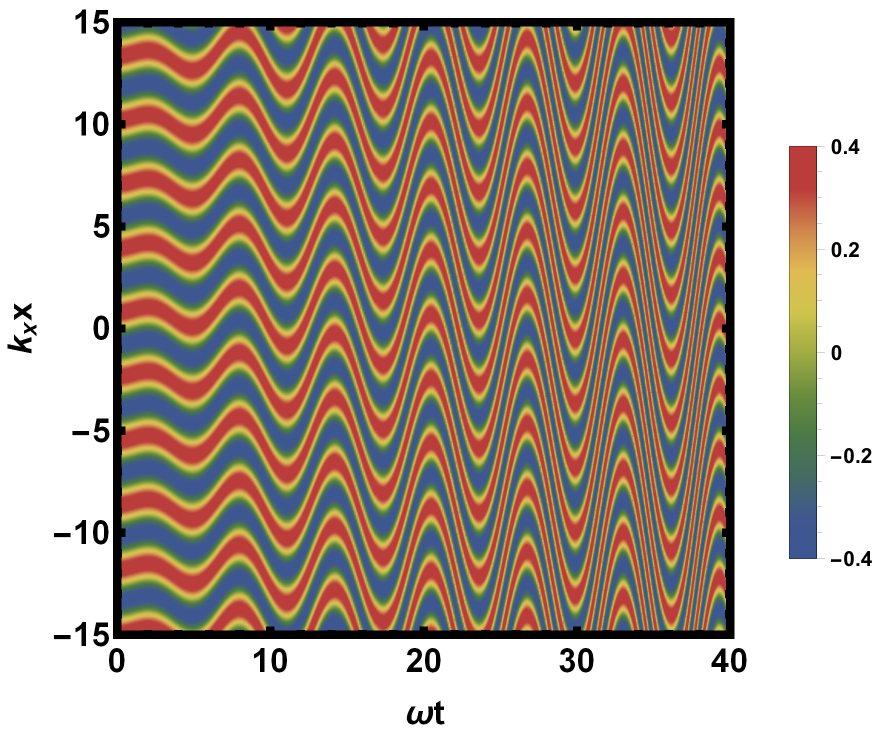}
		\label{figure0}}\\
	\subfloat[: $\delta L=0$]{\includegraphics[width=0.475\linewidth, height=0.08\textheight]{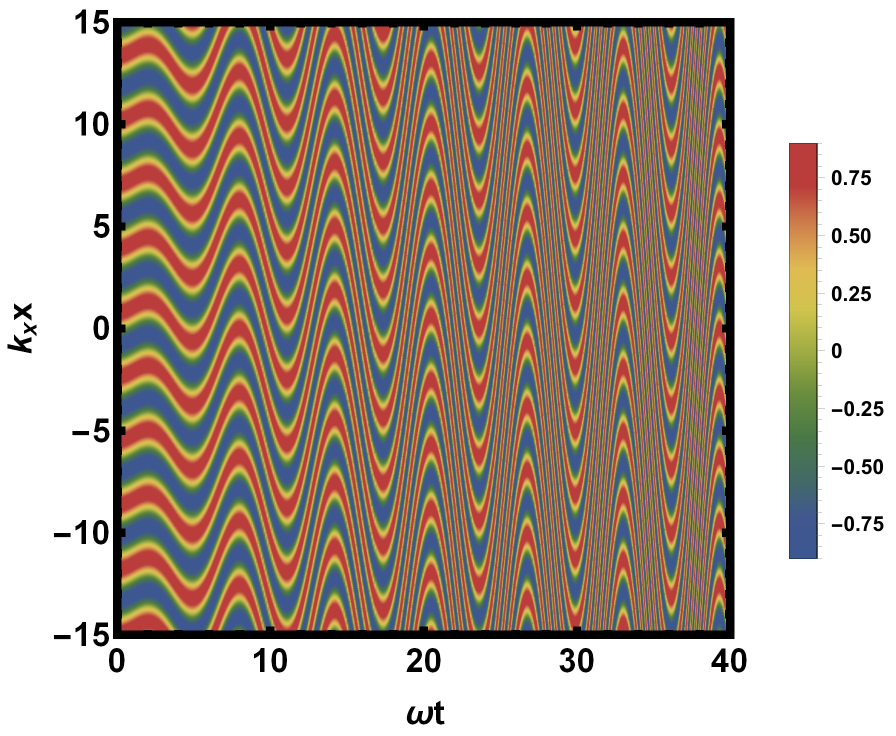}
		\label{figu10}}
	\subfloat[: $\delta L=0.8$]{\includegraphics[width=0.475\linewidth, height=0.08\textheight]{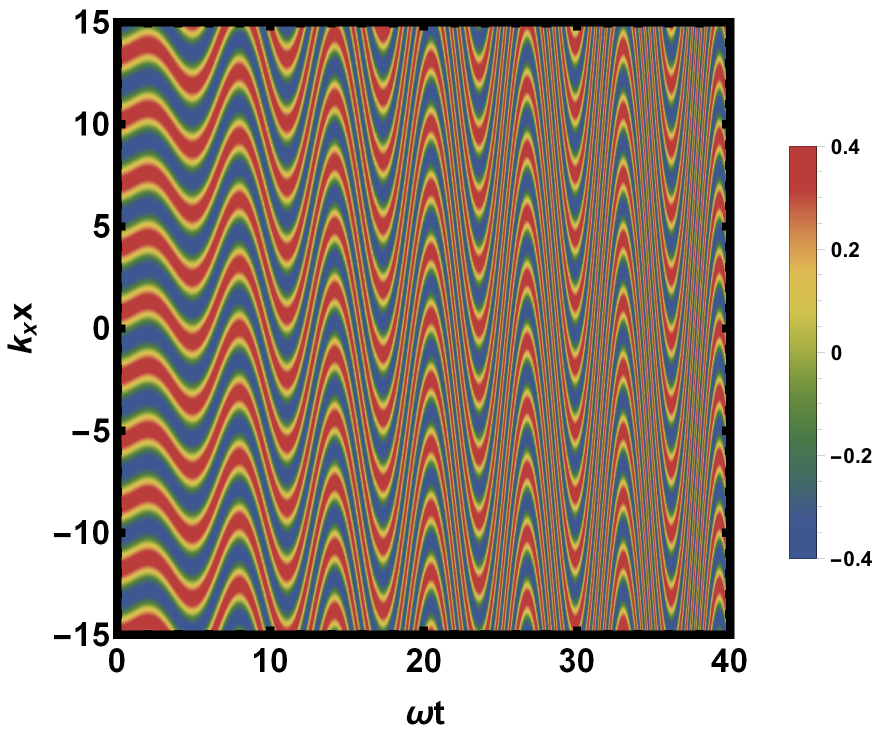}
		\label{figO}}
	\caption{{{\color{black}Current density $j_{0y}/2v_{F}$ versus time $\omega{t}$ and wave vector $k_xx$ for $\tau_{z}=1$, $x=\pi L/2$, $\omega=\pi v_{F} /2L$, $\varepsilon L=0.9$, $\alpha=\sqrt{{L}/{2}}$, $L=1$ $\text{nm}$. 
				(a)/(b){:} $U_{0}L/\hbar v_{F}=0.3$.   (c)/(d){:}  $U_{0}L/\hbar v_{F}=0.6$.}}}
	\label{figui}
\end{figure}
	
	{\color{black}
		
Fig.~\ref{figui} illustrates the normalized current density \( j_{0y}/2v_F \) as a function of the time \( \omega t \) and the wave vector \( k_x x \). The system parameters are chosen as	\( \omega = \pi v_F / 2L \), \( \tau_z = 1 \), \( x = \pi L/2 \), \( \varepsilon L = 0.9 \), \( \alpha = \sqrt{L/2} \), and \( L = 1\,\mathrm{nm} \). Figs.~\ref{figui}(a,b) correspond to the weak potential regime
\( U_0 L/\hbar v_F = 0.3 \), whereas Figs.~\ref{figui}(c,d) represent the stronger potential regime \( U_0 L/\hbar v_F = 0.6 \). For \( \delta L = 0 \), we notice that the current density exhibits regular oscillatory patterns characterized by quasi-periodic fringes in both time and momentum space \cite{25}. These oscillations originate from the coherent dynamics of Dirac fermions under periodic driving and from the interference between propagating modes.  When $\delta L = 0.8$, we clearly see that the interference pattern is significantly modified. The fringes become distorted and the maximum current amplitude is enhanced, particularly for intermediate values of \(k_x x \). A comparison between the two potential strengths shows that increasing \( U_0 L/\hbar v_F \) substantially amplifies the current oscillations. In the strong potential regime, the fringes become denser and the current amplitude increases, reflecting a stronger coupling between charge carriers and the external potential. This regime highlights an enhanced sensitivity of the transverse current density to the driving field parameters. Overall, these numerical results demonstrate that \( j_{0y}/2v_F \) is strongly modulated by both the energy gap \( \delta L \) and the strength of the applied potential \( U_0 L/\hbar v_F \). The interplay between these parameters offers an effective way to control the temporal and spectral characteristics of the current,  which can be exploited in Dirac-fermion–based electronic devices subjected to periodic driving.}
\begin{figure}[H]\centering
	\subfloat[: $k_{x}x=0$]{\includegraphics[width=0.475\linewidth, height=0.093\textheight]{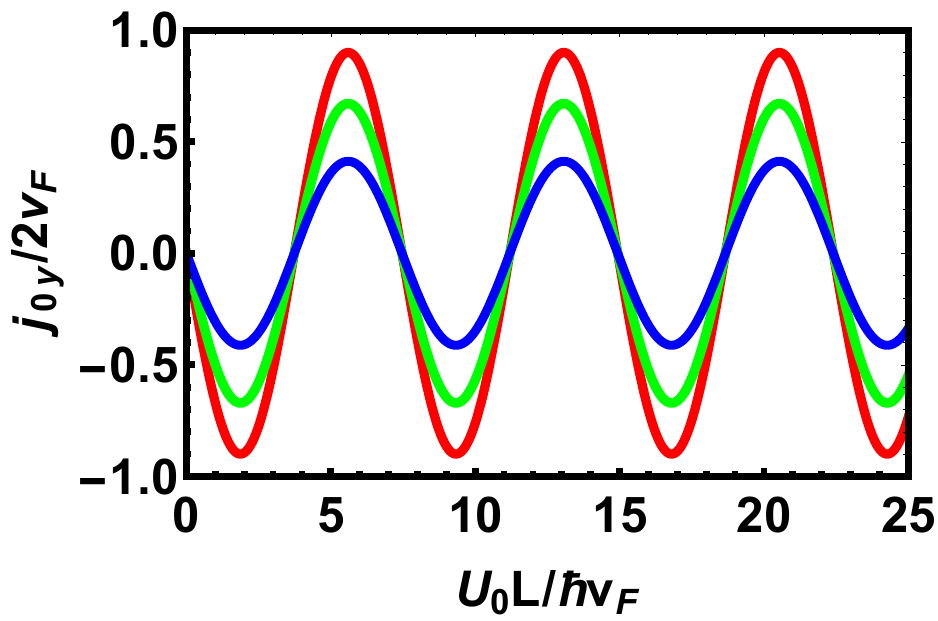}
		\label{figyr}}
	\subfloat[: {\color{black}$k_{x}x=-\pi/8$}]{\includegraphics[width=0.475\linewidth, height=0.093\textheight]{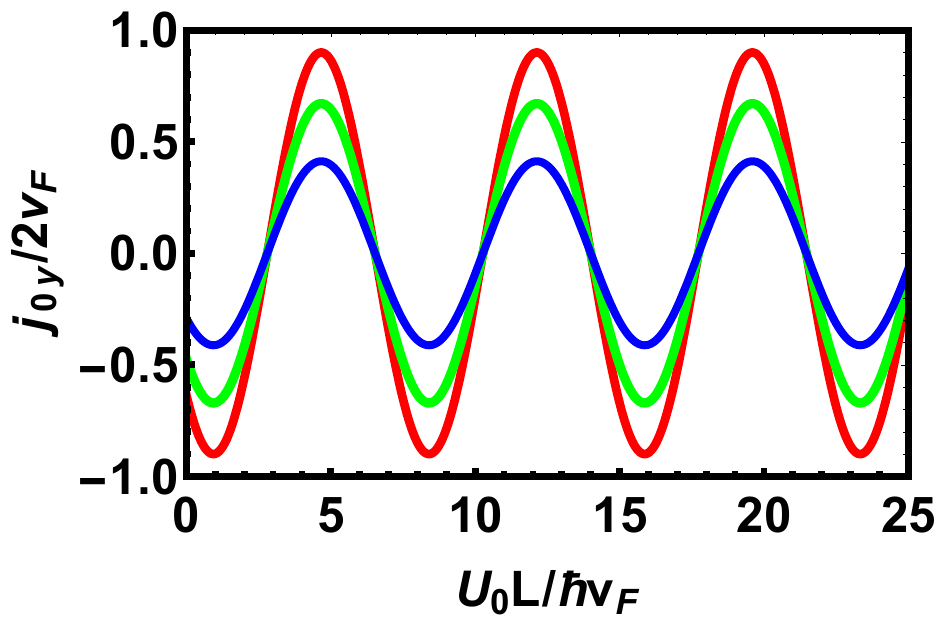}
		\label{figry1}}
	\caption{Current density $j_{0y}/2v_{F}$ versus the amplitude potential $U_{0}L/\hbar v_{F}$ for $\tau_{z}=1$, $x= \pi L/2$, $\omega= v_{F}/L$, $\varepsilon L=0.9$, $\alpha=\sqrt{{L}/{2}}$, $L=1$ $\text{nm}$, $\omega{t}=1$, $\delta L=0$ (red line), $0.6$ (green line), $0.8$ (blue line). (a){:} $k_{x}x=0$,  (b){:} {\color{black}$k_{x}x=-\pi/8$}.} 
	\label{fit}
\end{figure} 
	To highlight the effect of  the wave vector  on the current density in Fig. \ref{fit}, we present $j_{0y}/2v_{F}$ versus the potential amplitude $U_{0}L/\hbar v_{F}$ for $\tau_{z}=1$, $x=\pi L/2$, $\omega=v_{F}/L$, $\varepsilon L=0.9$, $\alpha=\sqrt{{L}/{2}}$, $L=1$ $\text{nm}$, with $\omega t=1$ and three gaps: $\delta L=0$ (red line), $0.6$ (green line), $0.8$ (blue line) with (a){:} $k_{x}x=0$, (b){:} {\color{black}$k_{x}x=-\pi/8$}. Note that these plots  display periodic behavior, characterized by a long temporal period.
	At spatial temporal matching resonance $\omega=v_{F}/L$ for $0\leq U_{0}L/\hbar v_{F}\leq 3.5$ and $k_{x}x=0$ in Fig. \text{\ref{fit}}(a), $j_{0y}/2v_{F}$ 
	{\color{black}exhibits an oscillatory behavior as the potential amplitude increases. The amplitude and sign of the current are strongly influenced by the value of $\delta L$. As $\delta L$ increases from $0$ to $0.8$,  the oscillations become more pronounced, indicating an enhanced sensitivity of the  current to the modulation strength. This behavior reflects the constructive and destructive interference between propagating states induced by the periodic potential.} However, for {\color{black}$k_{x}x=-\pi/8$} in Fig. \text{\ref{fit}}(b), 
	{\color{black}a noticeable phase shift in the oscillatory pattern is observed}, in contrary to the case $k_xx=0$. 	{\color{black}Although the overall dependence on  $U_{0}L/\hbar v_{F}$ remains oscillatory, the extrema of the current are shifted, and the magnitude of the current is modified. This demonstrates that $j_{0y}/2v_{F}$ is highly sensitive to the longitudinal wave-vector phase $k_{x}x$, highlighting the role of quantum phase effects in controlling the current response.}
	This result shows that the effect of the gap reduces the phenomenon of resonance when $\delta L<\varepsilon L$ and therefore the gap plays an important role in adjusting current density behavior.

	{

\section{Regime of Validity and Experimental Considerations}\label{V}

In the present analysis, we consider coherent quantum transport through a periodically driven region in graphene. Our approach implicitly assumes the ballistic regime, in which the elastic scattering time $\tau$ associated with disorder is the largest relevant timescale in the problem. Under this assumption, electrons can propagate across the driven region without experiencing significant scattering events, allowing the coherent dynamics responsible for the predicted current oscillations to develop. 
This regime can be expressed through the following conditions
\begin{align}
	\dfrac{E_F \tau}{\hbar} \gg 1, \quad \omega \tau \gg 1, \quad \dfrac{\Delta \tau}{\hbar} \gg 1 .
\end{align}
The first condition $E_F\tau/\hbar \gg 1$ establishes that quasiparticles, within the Fermi level range, maintain their distinct existence despite minor disorder. This condition corresponds to a weak-disorder metallic state, in which electronic states experience minimal widening due to scattering phenomena.
The second condition $\omega\tau\gg1$ means that the timescale of the external periodic driving is much smaller than the typical scattering time. In this regime, the system can respond coherently to the oscillating field and exhibit well-defined AC transport properties. Within the Drude framework, $\omega\tau\gg1$ is often used to identify the regime in which the electronic response remains phase coherent and frequency-dependent transport phenomena become observable \cite{AshcroftMermin1976}. In our context, this condition ensures that periodically driven Dirac fermions can undergo coherent oscillatory dynamics before scattering processes disrupt their motion.
The third condition, $\Delta\tau/\hbar \gg 1$, guarantees that the mass term $\Delta$, which opens a gap in the Dirac spectrum, stays as a valid energy scale that holds scientific value. The condition allows the spectral features of the gap to remain unchanged because disorder does not cause significant broadening, which enables clear observation of current changes that depend on the gap. Theoretical studies of graphene transport and Dirac materials use similar methods  \cite{CastroNeto2009}.

The experimental results support the above assumptions and match the behavior of graphene-based systems that can be tested in laboratories. In high-quality graphene devices, the elastic scattering times reach a maximum extending to the picosecond range, resulting in average free paths ranging from hundreds of nanometers to several micrometers. These conditions enable coherent quantum effects to dominate the behavior of the system, causing the system to naturally enter ballistic or quasi-ballistic transport mode. Modern graphene platforms with tunable band gaps allow researchers to adjust the parameters $E_F$, $\Delta$, and $\omega$ through experimental procedures representing real-world conditions. Under these specific conditions, the experimental evidence should demonstrate the resonant characteristics and Josephson-like current oscillations predicted by our model.

%

If the above criteria are not fully satisfied, disorder and thermal effects will gradually reduce the visibility of the predicted phenomena. For example, when $\omega\tau \lesssim 1$, scattering processes occur on time scales comparable to the external driving period. In this situation, the disorder-induced spectral broadening leads to two effects: the loss of sharp resonances during photon-assisted transport and the decrease of current oscillation magnitudes. Spectral features related to the mass gap begin to lose sharpness due to disorder-induced spectral widening when the condition $\Delta\tau/\hbar \lesssim 1$ occurs. 
The temperature of the system affects its behavior. At elevated temperatures, the electronic distribution expands around the Fermi level due to thermal fluctuations, which have two effects: smoothing out oscillating current patterns and dimming the strength of the resonance effect. Josephson-like current oscillations maintain their basic operational method as long as phase coherence exists across the entire driven area.
Our theoretical model is valid as long as these conditions are met. They suggest that the predicted effects are most clearly observable in high-mobility, low-disorder graphene samples operating at moderate temperatures, where coherent transport and well-defined quasiparticle dynamics can be sustained.
%
}

	\section{Conclusion}

	We explored the impact of an energy gap on Josephson-like currents in gapped graphene under the influence of a scalar potential with both spatial and temporal modulation, $U(x,t)$. Our analysis extends previous results reported in \cite{25}, originally derived for gapless graphene, to incorporate the effects of a finite bandgap. By solving the Dirac equation in the presence of this gap, we derived the energy spectrum and corresponding eigenspinors, which served as the foundation for calculating the current density components associated with Josephson-type effects. Two specific forms of external potentials were considered to illustrate the behavior of the system under different physical scenarios.
	We examined the behavior of the current density under various conditions, focusing particularly on the role of the energy gap $\delta$, the transverse momentum $k_y$, the amplitude of the driving potential $U_0$, and the wave vector $k_x$ along the propagation direction.

	For a temporally periodic potential at normal incidence (\( k_y = 0 \)) and under Shapiro resonance, 
	we found that the current density exhibits clean sinusoidal oscillations between positive and negative values 
	when the energy \( \varepsilon \) exceeds the gap \( \delta \). 
	{This behavior is reminiscent of a Josephson-like effect.} 
	But when the system is away from the Shapiro condition, the current becomes aperiodic 
	and decreases in amplitude as \( \delta L \) increases, 
	so there is a damping effect that suppresses the {Josephson-like response.}
	When finite transverse momentum is introduced ($k_y \neq 0$), a different class of oscillations emerges. These grow more pronounced as both the energy gap and time increase, pointing to a strong sensitivity of the current to valley and angular-dependent transport. Interestingly, the resonance phenomena persist in this regime regardless of the Shapiro step index $n$, highlighting the robustness of this effect across a wide range of system parameters.
	In the case of a combined spatially and temporally periodic potential, we demonstrated that the current density can be finely tuned to switch between positive and negative values by manipulating the gap, the driving strength $U_0$, and the sign and magnitude of $k_x$. This tunability shows the rich dynamics enabled by the material band structure and the external driving fields. 
	

    		Importantly, the predicted effects are experimentally accessible with currently available platforms and probes. 
    		Several graphene-based systems show the capability to adjust their band gap because graphene on h-BN substrates \cite{Hunt2013}, epitaxial graphene on SiC \cite{Zhou2007,Emtsev2008}, and electrically gated bilayer graphene \cite{Zhang2009,Castro2007} show band gap values that range from a few meV to hundreds of meV.
    		In these devices, a gate-defined barrier (or junction region) can be fabricated and driven periodically in the GHz--THz regime using microwave irradiation, ultrafast optical pumping, or on-chip high-frequency gating. This enables realistic experimental conditions for the detection of the predicted Josephson-like oscillations and photon-assisted transport features \cite{McIver2020}.
    		Experimentally, the key signatures discussed in our work can be tested through driven transport measurements by monitoring the dc and/or ac current response as a function of (i) the induced gap (tunable by gate bias in bilayer graphene or by substrate engineering), (ii) the driving frequency and amplitude, and (iii) the barrier parameters. 
    		Time-resolved and lock-in measurements can directly probe (1) the gap-dependent suppression or disappearance of Shapiro-like resonances, (2) irradiation-induced current sign changes, and (3) the emergence of transverse-momentum (and valley) dependent contributions, thereby providing a direct experimental verification of our main theoretical predictions \cite{Yoshikawa2017,Frisenda2018}.

\end{document}